\documentclass[twocolumn]{aastex62}

\graphicspath{{./}{figures/}}

\usepackage{color,subfigure}

\usepackage{ulem}
\usepackage{amsmath,bm}
\usepackage{commath}
\usepackage[figuresright]{rotating}
\usepackage{tabularx}
\usepackage[vlines]{tabularht}
\usepackage{multirow}
\usepackage{natbib}
\shorttitle{Photometric Quasars in DES}
\shortauthors{Yang \& Shen}
\begin{document}

\title{A Southern Photometric Quasar Catalog from the Dark Energy Survey Data Release 2}

\correspondingauthor{Qian Yang}
\email{qian.yang@cfa.harvard.edu} 

\author[0000-0002-6893-3742]{Qian Yang}
\affiliation{Department of Astronomy, University of Illinois at Urbana-Champaign, Urbana, IL 61801, USA}
\affiliation{National Center for Supercomputing Applications, University of Illinois at Urbana-Champaign, Urbana, IL 61801, USA}

\author[0000-0003-1659-7035]{Yue Shen}
\affiliation{Department of Astronomy, University of Illinois at Urbana-Champaign, Urbana, IL 61801, USA}
\affiliation{National Center for Supercomputing Applications, University of Illinois at Urbana-Champaign, Urbana, IL 61801, USA}

\begin{abstract}
We present a catalog of 1.4 million photometrically-selected quasar candidates in the southern hemisphere over the $\sim 5000\,{\rm deg^2}$ Dark Energy Survey (DES) wide survey area. We combine optical photometry from the DES second data release (DR2) with available near-infrared (NIR) and the all-sky unWISE mid-infrared photometry in the selection. We build models of quasars, galaxies, and stars with multivariate Skew-t distributions in the multi-dimensional space of relative fluxes as functions of redshift (or color for stars) and magnitude. Our selection algorithm assigns probabilities for quasars, galaxies, and stars, and simultaneously calculates photometric redshifts (photo-$z$) for quasar and galaxy candidates. Benchmarking on spectroscopically confirmed objects, we successfully classify (with photometry) 94.7\% of quasars, 99.3\% of galaxies, and 96.3\% of stars when all IR bands (NIR $YJHK$ and WISE $W1W2$) are available. The classification and photo-$z$ regression success rates decrease when fewer bands are available. Our quasar (galaxy) photo-$z$ quality, defined as the fraction of objects with the difference between the photo-$z$ $z_p$ and the spectroscopic redshift $z_s$, $|\Delta z| \equiv |z_s - z_p|/(1 + z_s)\le 0.1$, is 92.2\% (98.1\%) when all IR bands are available, decreasing to 72.2\% (90.0\%) using optical DES data only. Our photometric quasar catalog achieves estimated completeness of 89\% and purity of 79\% at $r<21.5$ (0.68 million quasar candidates), with reduced completeness and purity at $21.5<r\lesssim 24$. Among the 1.4 million quasar candidates, 87,857 have existing spectra and 84,978 (96.7\%) of them are spectroscopically confirmed quasars. Finally, we provide quasar, galaxy, and star probabilities for all (0.69 billion) photometric sources in the DES DR2 coadded photometric catalog. 
\end{abstract}


\section{Introduction} \label{sec:introduction}

Active galactic nuclei (AGNs) and their high-luminosity counterparts, quasars, are accreting supermassive black holes (SMBHs) at the center of massive galaxies. Understanding the evolution of the quasar population across cosmic time is crucial to understanding the physics of accretion and the coevolution of SMBHs and their host galaxies.

Large quasar surveys provide the necessary samples for measuring the abundance of quasars as functions of redshift and luminosity. In addition, these surveys enable a broad range of quasar science, such as quasar lens searches and their constraints on cosmology and the evolution of massive galaxies \citep[e.g.,][]{Oguri2006,Oguri2012}, finding projected quasar pairs \citep[e.g.,][]{Hennawi2006, Prochaska2013} and binary quasars \citep[e.g.,][]{Hennawi2006b, Hennawi2010}, and measuring quasar clustering \citep[e.g.,][]{Martini2001, Shen2007}. Large quasar surveys also provide an opportunity to identify rare objects, such as extreme variability quasars \citep[e.g.,][]{Rumbaugh_etal_2018}, and study their fueling mechanisms \citep[e.g.,][]{MacLeod2016, Yang_etal_2018}. Finally, large samples of quasars are often used to define the celestial reference frame \citep[e.g.,][]{Gaia2018b}.

Distant quasars have been discovered beyond redshift 7 \citep{Mortlock2011, Banados2018, Yang2020, Wang2021}, where massive SMBHs had formed within less than one billion years after the Big Bang. A large sample of quasars over a broad range of redshifts enables the study of the evolution of SMBHs as well as the intergalactic medium. For example, the Ly$\alpha$ forest in quasar spectra can be used to measure baryon acoustic oscillations as a probe for cosmology \citep{Dawson_etal_2013}.

While the Sloan Digital Sky Survey \citep[][]{York2000} has provided large samples of quasars in the northern hemisphere, there is a lack of large spectroscopically confirmed quasar samples in the southern hemisphere. There are over 750k quasars in the SDSS DR16 quasar catalog \citep{Lyke2020}. In contrast, there are less than 24k spectroscopically confirmed quasars in the southern hemisphere in the Million Quasars Catalog \citep[][v6.5]{Flesch2019} with $\rm{Decl.} < -10^\circ$. A large sample of quasars in the southern hemisphere will be important for quasar-related studies in the next few decades, given increasing investments of ground-based facilities covering the southern sky, in particular, the Vera C. Rubin Observatory Legacy Survey of Space and Time \citep[LSST,][]{LSST_ref}. 

The Dark Energy Survey \citep[DES;][]{Abbott2018} is a wide-area visible and near-infrared imaging survey covering $\sim$5000 deg$^2$ of the high Galactic latitude sky, mostly in the southern hemisphere. The multi-band deep DES photometry enables the photometric selection of a large quasar sample in the southern hemisphere. In this work, we perform a systematic selection of quasar candidates using public photometric data from DES over the $\sim$5000 deg$^2$ wide-survey footprint, combined with publicly available near-infrared (NIR) and mid-infrared (MIR) photometric data. With these data, we classify quasars, galaxies, and stars in all DES detected photometric sources with probabilities and estimated photo-zs (for galaxy and quasar candidates). 

The structure of this paper is as follows. In \S\ref{sec:data}, we describe the imaging surveys and the training samples. We describe the selection methods in \S\ref{sec:method}. We present the quasar catalog in \S\ref{sec:results}, and discuss the selection completeness and efficiency in \S\ref{sec:disc}. We conclude in \S\ref{sec:summary}. Throughout this paper, we adopt a flat $\Lambda$CDM cosmology with parameters $\Omega_{\Lambda}=0.7$, $\Omega_{\rm m}=0.3$, and $H_0=70$ km s$^{-1}$ Mpc$^{-1}$. The Milky Way extinctions of extragalactic objects in DES bands are corrected using the dust reddening map of \citet{Schlegel1998}. In this work, the term ``quasar'' is used to broadly refer to an unobscured broad-line AGN regardless of its luminosity. We also only consider quasar targets where the continuum emission is dominated by the quasar rather than the host galaxy.

\section{Data and Samples} \label{sec:data}

\begin{deluxetable*}{lccll}
\tablecaption{Photometric Survey \label{tab:survey}}
\tablewidth{1pt}
\tablehead{
\colhead{Survey} &
\colhead{Data Release} &
\colhead{Area (deg$^2$)} &
\colhead{Filter} &
\colhead{Depth (AB, 5$\sigma$)}
}
\startdata
DES & DR2 & 5000 & $g, r, i, z, Y$ & 25.0, 24.5, 23.7, 22.6, 21.3 \\
VIKING & DR5 & 1500 & $Y, J, H, K_s$ & 21.9, 21.8, 21.2, 21.1 \\
VHS    & DR6 & 8300 & $Y, J, H, K_s$ & 20.7, 21.0, 20.5, 20.2 \\
ULAS    & DR11 & 4000 & $Y, J, H, K$ & 21.0, 20.7, 20.3, 20.2 \\ 
UHS & DR1 & 12700  & $J$ & 20.3 \\ 
2MASS & $\cdots$ & All sky & $J, H, K_s$ & 18.0, 17.6, 17.4 \\
unWISE & NEO6 & All sky & $W1, W2$ & 21.7, 20.9 \\
Gaia & DR2 & All sky & $\cdots$ & $\cdots$ \\
\enddata
\tablecomments{NEO6: Up to year 6 of NeoWISE-Reactivation. We only use astrometry information from Gaia DR2.
}
\end{deluxetable*}

\subsection{Imaging Surveys \label{sec:survey}}
We use the second public data release (DR2) of the DES  \citep[][]{Abbott2021}, including data from the DES wide-area survey covering $\sim$ 5000 deg$^2$ of the southern Galactic cap in five broad photometric bands ($grizY$). We use the DES DR2 coadded photometric catalog, including $\sim$691 million distinct astronomical objects, the vast majority of which are non-transient and non-moving objects. For the DES coadded photometry, we use the IMAFLAGS\_ISO flag to remove unreliable detections, which is set if any pixel is masked in all of the contributing exposures for a give band. This flag is mainly set for saturated objects and objects with missing data \citep{Abbott2018}. The median coadded catalog point-source depth at ${\rm S/N} = 5$ in the $grizY$ bands are 25.0, 24.5, 23.7, 22.6, and 21.3, respectively (PSF mag). We use both PSF and AUTO photometry in DES depending on the fitting template class (see below).

\begin{figure*}[!ht]
\centering
\includegraphics[width=0.95\textwidth]{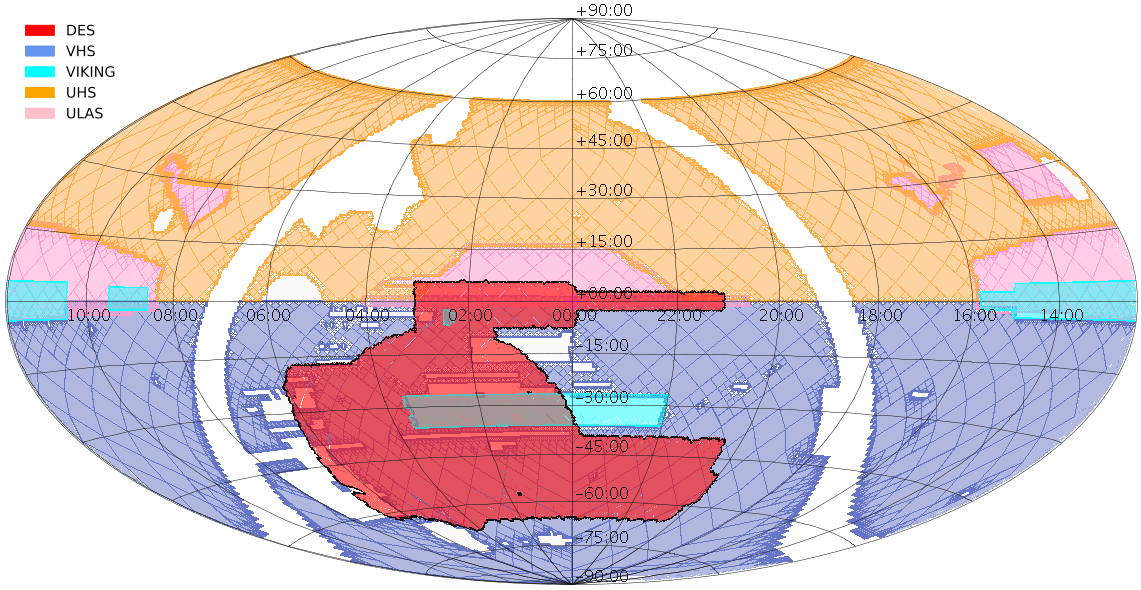}
\caption{Sky coverage of DES and NIR surveys in equatorial coordinates. WISE, 2MASS, and Gaia are all-sky surveys and not shown here. DES: shaded red area with black outline; VHS: shaded blue area with meshes; VIKING: shaded cyan area; UHS: shaded orange area with meshes; ULAS: shaded pink area with meshes. These NIR surveys (except for 2MASS) barely overlap with each other. For areas not covered by VHS, VIKING, UHS, or ULAS, we used 2MASS NIR data.
\label{fig:NIR}}
\end{figure*}

For the NIR data, we make use of all public NIR imaging in the DES area, including the VISTA Hemisphere Survey \citep[VHS,][]{McMahon2013}, the VISTA Kilo-Degree Infrared Galaxy Survey \citep[VIKING,][]{Edge2013}, the UKIDSS Large Area Surveys \citep[ULAS,][]{Lawrence_etal_2007}, and the Ukirt Hemisphere Survey \citep[UHS,][]{Dye2018}. For these NIR surveys, we use the aperture-corrected magnitude in a 2$\arcsec$ diameter circle. For areas not covered by these NIR surveys, we use the shallower all-sky Two Micron All Sky Survey \citep[2MASS,][]{Skrutskie_etal_2006} data\footnote{https://doi.org/10.26131/IRSA2}. Figure \ref{fig:NIR} shows the sky coverages of different NIR imaging surveys. The VHS survey covers most of the DES area in the southern sky. We summarize the depths of these NIR surveys in Table \ref{tab:survey}. We ignore the slight filter differences among different NIR surveys, which result in minor magnitude differences (normally less than 0.05 mag). 

In the MIR, we use the unblurred coadds of the Wide-field Infrared Survey Explorer \citep[WISE,][]{Wright_etal_2010} imaging data \citep[unWISE,][]{Lang_2014, Meisner2019}. We use the unWISE photometry from coadds of the WISE and NEOWISE (through the sixth year NEOWISE data release in 2020). The unWISE catalog has advantages over the AllWISE catalog\footnote{https://doi.org/10.26131/IRSA1} since it is based on significantly deeper imaging and features improved photometric modeling in crowded regions \citep{Schlafly2019}. The 5$\sigma$ depth in AB magnitude in unWISE $W1$ and $W2$ bands are 21.7 and 20.9, respectively.

Gaia DR2 \citep{Gaia2018} contains celestial positions for 1.7 billion sources, and parallaxes and proper motions for 1.3 billion sources. We use Gaia astrometry information to help rule out stars with detected proper motion or parallax.

\subsection{Training Samples}\label{sec:traning}

We consider three object classes, quasars, stars, and galaxies, for which we build empirical color templates from training samples. Stars and most quasars are point-like objects, and galaxies are mostly extended sources. Each object is fit to three classes of color templates (quasar, star and galaxy). When fitting with quasar and star templates, we default to use DES PSF photometry for the object. When fitting with the galaxy template, we default to use the AUTO photometry in DES. At the faint end, for some objects without DES PSF photometry in some bands, we use the DES AUTO photometry for all three classes.

We then use spectroscopically confirmed quasars, stars and galaxies to build our color templates. For quasars, we start from the SDSS DR16 quasar catalog \citep{Lyke2020}, but remove unreliable high-redshift quasars misclassified by the SDSS pipeline. Specifically, we removed $z>5$ quasars that were only classified as ``QSO'' by the pipeline but not confirmed by visual inspection (most of these are pipeline misclassifications of low-redshift quasars or non-quasars). Next, we supplement spectroscopically confirmed quasars from the Million Quasars Catalog, v6.5 \citep{Flesch2019}. We added sources with type as ``Q" and ``A", which are broad-line quasars and broad-line Seyferts respectively. This supplementary sample is necessary because it includes confirmed high-redshift quasars at $z>5$; and quasars from the 2dF QSO Redshift Survey \citep[2QZ,][]{Croom2004}, the 2dF-SDSS LRG and QSO survey \citep[2SLAQ,][]{Croom_etal_2009}, the Australian Dark Energy Survey \citep[OzDES,][]{Yuan2015}, and the Large Sky Area Multi-object Fiber Spectroscopic Telescope (LAMOST) quasar catalog \citep{Yao2019}. The redshifts of the majority of the spectroscopically confirmed quasars are lower than 3.5 (99\% of SDSS quasars). The number of spectroscopically confirmed quasars decreases rapidly with redshift, specifically from 9178 at $3 < z < 3.1$ to 634 at $4 < z < 4.1$, and to 37 at $6 < z < 6.1$. So at the high redshift end, using only these confirmed quasars may lead to strong biases from individual quasars. To improve the color coverage of $z>3.5$ quasars, we add simulated quasars \citep{McGreer_etal_2013} at high redshift ($z=3.5-8$)\footnote{The DECam Y band extends to $\sim$10700 $\rm{\AA}$. Therefore, at $z>7.8$, the Ly$\alpha$ emission of quasars starts to drop out of DES $Y$ band.}. The simulated quasar models include a broken power-law continuum, UV/optical emission lines, pseudo-continuum from FeII emission, and redshift-dependent Ly$\alpha$ forest absorption due to neutral hydrogen. The numbers per redshift bin of simulated $z>3.5$ quasars are close to those of SDSS quasars at $1.5<z<3.5$. We simulated a large number of quasars to ensure a sufficient statistical sample to avoid the impact of random fluctuations.

We consider contamination from stars and galaxies in our quasar selection. We use spectroscopic galaxies and stars from the SDSS DR16 \citep{Ahumada2020}. SDSS galaxies are representative of the low-redshift galaxy population but are limited to $z\lesssim 1$ given the nature of optical SDSS surveys. However, the lack of representation of $z\gtrsim 1$ galaxies in the training sample does not affect our quasar selection, since these high-$z$ galaxies are typically much fainter in the observed-frame optical than our quasar targets. We supplement the sample with stars from the fifth data release of the LAMOST survey \citep{Luo2015}. We restrict to high-galactic-latitude stars in LAMOST with $|b|>20^\circ$, as the DES footprint is all at $|b|>20^\circ$. Compared with SDSS, the supplemental LAMOST stars are mainly at the bright end ($i<18$). The star training sample is representative of different types of stars, from white dwarf to late-type stars. For example, more than half of the 68k white dwarfs from the Montreal White Dwarf Database\footnote{http://www.montrealwhitedwarfdatabase.org} are in our star training sample, and the other half are mainly out of the SDSS sky coverage. Among the 10k brown dwarfs compiled by \citet{Best2018} from the DwarfArchives\footnote{http://DwarfArchives.org.}, 83\% of them are in our star training sample.

We summarize the numbers of different classes of objects from different catalogs in Table \ref{tab:number}. We cross-matched the sources with the imaging surveys described in \S\ref{sec:survey} with a search radius of 2$\arcsec$. The numbers of sources detected by different imaging surveys are also included in Table \ref{tab:number}. Since most SDSS sources are in the northern sky and not covered by DES, we convolve the SDSS spectra with the DES filter curves to generate synthetic DES photometry in the training samples. Because the DES $Y$ band spans from $\sim$9400$\rm{\AA}$ to $\sim$10700$\rm{\AA}$, we do not use spectra taken by the SDSS-I/II spectrographs (only up to 9200$\rm{\AA}$), and use spectra taken by the SDSS BOSS spectrographs (up to 10400 $\rm{\AA}$) whenever applicable.

\begin{table*}
\centering
\caption{Spectroscopically-confirmed Training Samples \label{tab:number}}
\begin{tabular}{l|r|rrr|r|r|rr}
\hline \hline
Class & \multicolumn{4}{c|}{QSO} & \multicolumn{1}{c|}{Galaxy} & \multicolumn{3}{c}{Star} \\
\hline
Database & All & SDSS & Milliquas & Simulation & All (SDSS) & All & SDSS & LAMOST \\
\hline
Sample Size & 1,252,844 & 744,293 & 63,948 & 444,603 & 2,655,705 & 4,426,020 & 1,028,777 & 3,397,243 \\
\hline
DES & 102,321 & 85,181 & 17,140 & $\cdots$ & 318,438 & 160,165 & 40,141 & 120,024 \\
DES (filter) & 1,023,645 & 579,042 & $\cdots$ & 444,603 & 1,520,671 & 426,334 & 426,334 & $\cdots$ \\
\hline
unWISE & 1,239,666 & 733,102 & 61,961 & 444,603 & 2,614,053 & 4,277,893 & 914,133 & 3,363,760 \\
\hline
Gaia  & 518,584 & 468,385 & 50,199 & $\cdots$ & 335,923 & 4,185,994 & 819,135 & 3,366,859 \\
\hline
NIR (Y)  & 631,885 & 155,399 & 31,883 & 444,603 & 857,511 & 1,200,916 & 276,327 & 924,589 \\
NIR (J) & 859,988 & 367,988 & 47,397 & 444,603 & 2,068,340 & 4,104,015 & 731,234 & 3,372,781 \\
NIR (H) & 628,567 & 148,225 & 35,739 & 444,603 & 1,402,542 & 3,850,179 & 506,919 & 3,343,260 \\
NIR (K) & 646,936 & 163,652 & 38,681 & 444,603 & 1,430,952 & 3,766,737 & 469,049 & 3,297,688 \\
\hline
\end{tabular}
\tablecomments{We use spectroscopically-confirmed quasars/galaxies/stars from SDSS, quasars in the Million Quasars Catalog (Milliquas), and stars from LAMOST as our training samples. We cross-match the spectroscopically-confirmed samples with source catalogs from various imaging surveys, including DES, unWISE, Gaia, and NIR surveys (described in \S\ref{sec:survey}).
Since most SDSS sources are in the northern sky and not covered by DES, we convolve the SDSS spectra with the DES filter curves to generate synthetic DES photometry.
To improve the color coverage of high-redshift quasars, we add simulated high-redshift quasars (see \S\ref{sec:traning}).
}
\label{table:1}
\end{table*}

\section{Target Selection Algorithms} \label{sec:method}

\subsection{General Considerations}

The photo-$z$ problem is a regression problem, relying on the description of the probability distribution of redshift for a specific class of objects. Quasar target selection is a classification problem, depending on the probability estimates for different classes of objects, such as quasars, stars, and galaxies.

We briefly describe the prior, likelihood, and posterior probabilities in our Bayesian analysis. In Bayes' theorem, the posterior probability of model parameters $\theta$ given data $x$ can be written as\\
\begin{equation} \label{eq:posterior}
    p(\theta|\bm{x}) = \frac{p(\bm{x}|\theta)p(\theta)}{p(\bm{x})}\propto \mathcal{L}(\theta)p(\theta)
\end{equation}
where $\mathcal{L}(\theta) = p(\bm{x}|\theta)$ is the likelihood and $p(\theta)$ is the prior probability of model parameters $\theta$. Here $p(\bm{x})$ is the normalizing constant (also called evidence), and is usually ignored in the inference.

In our photo-$z$ regression problem, $\theta$ is the redshift, $\bm{x}$ is the multi-dimensional relative fluxes (i.e., flux ratio wrt the flux in a reference band). The prior distribution, $N(z)$, is the predicted number distribution of an object class (i.e., quasar or galaxy), as a function of redshift $z$.
We describe the prior distribution in \S\ref{sec:prior} and likelihood distribution in \S\ref{sec:likelihood}.

\subsection{Prior Distribution \label{sec:prior}}
The number densities of quasars and galaxies depend on redshift and luminosity, which can be estimated using the observed luminosity functions of quasars and galaxies. The number density of stars depends on stellar type and luminosity, as well as locations in the sky. All these number densities refer to the absolute sky densities of the three classes of objects. 

We implement the optical quasar luminosity function (QLF) from \citet{NPD2016}, which is based on quasars over a wide redshift range of $0.68<z<4.0$ and magnitudes as faint as 22.5 mag in $g$ band. We extrapolate this QLF to the faint end and the high redshift end. We also tested the bolometric QLFs from \citet{Hopkins_etal_2007} and \citet{Shen2020}. However, there are additional issues of utilizing these bolometric QLFs due to uncertainties in bolometric corrections and k-corrections, which are redshift and luminosity dependent. The QLF from \citet{NPD2016} works well for quasar photo-$z$ estimation over broad ranges of redshift and optical magnitude \citep{Yang2017}, as desired here.

We implement the galaxy luminosity function (GLF) from \citet{Montero-Dorta2009} based on SDSS data. There are different subclasses of galaxies, such as late-type and early-type galaxies. Our galaxy training sample is not rigorously labelled with different sub-types, lacking information such as star formation rate or morphology. So we simply treat all galaxies as a single class in this work. Most of these SDSS galaxies are at $z<1$, with a small fraction of them at higher redshifts. Since there are very few spectroscopically confirmed galaxies at $z \geq 1.5$ ($<$0.02\%) in the training sample, we restrict to $z < 1.5$. Galaxies at $z>1.5$ are too faint in the observed-frame optical to contaminate our quasar selection.

We estimate the number density of stars for typical high Galactic-latitude fields\footnote{In principle, our algorithm can implement a stellar number density as a function of Galactic latitude to improve the quasar selection in regions with higher stellar densities. For simplicity, in this work we only consider the typical stellar density at high Galactic latitudes, and leave this additional implementation to future work.}, from a Milky Way synthetic simulation with the Besan\c{c}on model \citep{Robin2003}. \citet{Yang2017} showed that the simulated star number distribution is close to the observed distribution. We performed such a simulation of stars with DES filters in a 100 deg$^2$ region with a central position at R.A.$ = 2$h and decl.$=-36^{\circ}$, which is close to the median central position of the DES survey. The number density of stars also depends on stellar types. Our star training sample is not well classified into different stellar types. Instead, we use color, $c$, as an alternative parameter for different stellar types. We describe what colors are used specifically for stars in \S\ref{sec:likelihood}.

Using the QLF, GLF, and star simulations described above, we derive the quasar number (per deg$^2$) prior distribution as a function of redshift, ${N_{\rm QSO}(z)}$; the galaxy number prior distribution as a function of redshift, ${N_{\rm Galaxy}(z)}$; and the star number prior distribution as a function of color, ${N_{\rm Star}(c)}$, in a set of magnitude bins. Our algorithm can be improved with better QLF and GLF for a wider range of redshifts and magnitudes. Our galaxy photo-$z$ can be further improved with galaxy training samples and GLFs labelled with different sub-types. 

\subsection{Likelihood Function}\label{sec:likelihood} 
The key problem of our target selection/classification is to describe the likelihood of a series of attributes, $\bm{x}$, for a given redshift and magnitude. Specifically, in our algorithm $\bm{x}$ represents the multi-dimensional relative fluxes. 

The colors of quasars change as a function of redshift, due to the shift of quasar emission lines moving in and out of different filters. Quasar colors also change as a function of magnitude due to multiple reasons: (1) the colors of quasar at the faint end or at low redshift are more affected by their host galaxy light; (2) quasars are usually bluer when brighter; (3) the equivalent widths of quasar emission lines are often anti-correlated with the continuum emission \citep[i.e., the Baldwin effect;][]{Baldwin1997}.

The colors of quasars at similar redshift and magnitude are usually similar. To fit the color distribution in multi-dimensional space, we can, for example, (1) fit the color in each color dimension with a Gaussian distribution, such as using the $\chi^2$ method \citep[e.g.,][]{Richards2001}; (2) fit the colors in multi-dimensional space with multivariate Gaussian distribution, such as using the multivariate $\chi^2$ method \citep[e.g.,][]{Weinstein2004}; (3) fit the colors in multi-dimensional space with mixture of multiple multivariate Gaussian distributions, such as the XDQSOz technique \citep{Bovy2011, Bovy2012}; (4) fit the colors in multi-dimensional space with Machine Learning techniques \citep[e.g.,][]{Yeche2010, Shu_etal_2019}; or (5) fit the colors in multi-dimensional space with more flexible parametric distribution, such as the multivariate Skew-t distribution \citep{Yang2017}.

Skewt-QSO is an algorithm for quasar selection and photo-$z$ estimation \citep{Yang2017}. The color distribution of quasars shows skewed and tail features mainly due to intrinsic dust reddening. Skewt-QSO describes the color distribution of quasars in a specific redshift and magnitude range by multivariate Skew-t distributions. \citet{Yang2017} demonstrated that the Skew-t distribution better describes the color distribution of quasars than Gaussian or Skew-normal distributions. Skewt-QSO also achieves better photo-$z$ accuracy compared to other quasar photo-$z$ algorithms, such as the empirical color–redshift relation \citep[e.g.,][]{Richards2001, Weinstein2004} and the XDQSOz algorithm \citep{Bovy2012}. Here we briefly describe the Skewt-QSO algorithm (see more details in \citealt{Yang2017}).

The probability density function (PDF) of a multivariate Skew-t distribution, denoted by $ST_n(\bm{\mu}, \bm{\Sigma}, \bm{\lambda}, \nu)$, can be expressed as \citep{Lachos2014},
\begin{equation}\label{eq:skewt}
  2~t_n(\bm{x}|\bm{\mu}, \bm{\Sigma}, \nu) ~ T\begin{pmatrix} \frac{\sqrt{\nu+n} ~ \bm{\lambda}^\top \bm{\Sigma}^{-1/2} (\bm{x}-\bm{\mu}) }{\sqrt{\nu + d}} | 0, 1, \nu + n \end{pmatrix}\ ,
\end{equation}
where $\bm{x}$ is the $n$-dimensional variate (relative fluxes), $\bm{\mu}$ is the mean vector,  $\bm{\Sigma}$ is the covariance matrix, $\nu$ is the degree of freedom, $\bm\lambda$ is the shape parameter, and $d$ is the Mahalanobis distance $d = (\bm{x} - \bm{\mu})^\top \bm{\Sigma}^{-1}(\bm{x} - \bm{\mu})$.
$t_n$ and $T$ denote the PDF and cumulative distribution function (CDF) of the Student-t distribution,
\begin{equation} \label{eq:student}
  \frac{\Gamma(\frac{n+\nu}{2})}{\Gamma(\frac{\nu}{2})(\nu\pi)^{n/2}}|\bm{\Sigma}|^{-1/2}\begin{pmatrix}1+\frac{d}{\nu}\end{pmatrix}^{-(\frac{n+\nu}{2})}
\end{equation}
where $\Gamma$ is the gamma function. When $\bm{\lambda}=0$ and $\nu = \infty$, the Skew-t distribution becomes the normal distribution, $N_n({\bm{\mu}, \bm{\Sigma}})$.

As redshift increases, the Ly$\alpha$ emission begins to drop out and the Lyman-$\alpha$ forest begins to move into blue bands. We use the relative fluxes instead of colors because at the faint end even negative flux (e.g., non-detection) is useful. \citet{Yang2017} used $r$ band as the reference band. However, the Ly$\alpha$ emission of $z>5$ quasars begins to drop out of $r$ band. Using a fixed reference band for relative fluxes will lead to large uncertainties for high-redshift quasars. Here we use a flexible reference band to compute relative fluxes. We choose the reference band as the band with the maximum signal-to-noise ratio (S/N) in DES $grizY$ photometry. 

The likelihood function of the multivariate attribute, $\bm{x}$, for a given $\theta$ can be described by the multivariate Skew-t distribution. Here, $\bm{x}$ is the multi-dimensional relative fluxes. For quasars and galaxies, $\theta$ is the redshift, $z$; for stars, $\theta$ is the color, $c$.

To model the colors of quasars (construct the likelihood functions), we divide the quasar training sample described in \S\ref{sec:traning} in redshift bins of $\Delta z = 0.05$ and magnitude bins of $\Delta m = 0.1$. This bin size is large enough to enclose enough quasars in one bin and small enough for quasar photo-$z$ estimation \citep{Yang2017}. To model the colors of galaxies, we divide the galaxy training sample in redshift bins of $\Delta z = 0.01$ and magnitude bins of $\Delta m = 0.1$. 
We divide the star training sample into color bins, where the color (wrt the reference band) can be treated as a parameter similar to redshift for quasars/galaxies. We use different colors (specifically $g-r$, $r-i$, $r-i$, $i-z$, or $z-Y$) when the reference band is different ($g$, $r$, $i$, $z$, or $Y$). We divide the star training sample into color bins of $\Delta c = 0.02$ and magnitude bins of $\Delta m = 0.1$.
Thus we obtain a series of Skew-t parameters $(\bm{\mu}, \bm{\Sigma}, \bm{\lambda}, \nu)$ as a function of redshift (or color) and magnitude for quasars, galaxies, and stars, respectively. Using the multivariate Skew-t distributions with these parameters, we obtain the likelihood functions to describe quasars, galaxies, and stars in the multi-dimensional color space as a function of redshift (or color), in different magnitude bins. 

For an object, at a given magnitude, with the multivariate attribute $\bm{x}$ (multi-dimensional relative fluxes) and their uncertainties, we using Equation (\ref{eq:skewt}) to estimate the likelihood in each quasar redshift bin, $\mathcal{L}_{\rm QSO}(z)$, the likelihood in each galaxy redshift bin, $\mathcal{L}_{\rm Galaxy}(z)$, and the likelihood in each star color bin, $\mathcal{L}_{\rm Star}(c)$.

\subsection{Joint Posterior Probability}
For an object with available photometric data in multiple bands (thus we know its magnitude and multi-dimensional relative fluxes), we obtain the joint posterior probability (Equation \ref{eq:posterior}) by combining the prior probability described in \S\ref{sec:prior} and the likelihood function described in \S\ref{sec:likelihood} for quasar, galaxy, and star classes, respectively. 

For the quasar class, we obtain the joint posterior probability at each redshift. The quasar class PDF is obtained as
\begin{equation} \label{eq:4}
  p_{\rm QSO}(z) = \mathcal{L}_{\rm QSO}(z) N_{\rm QSO}(z).
\end{equation}

We identify peaks in the PDF automatically using the $findpeaks$ function in R $pracma$ package\footnote{https://cran.r-project.org/web/packages/pracma/index.html}. We obtain the quasar photo-$z$, denoted as photoz-QSO, from the primary peak with the highest integrated probability within a redshift range $(z1, z2)$, where $z1$ and $z2$ denote the locations of zero probability on both sides of the peak as identified by {\it pracma}. A parameter $P_{\rm QSOz}$ describing the probability that the true redshift is located within the primary peak, $(z1_{\rm QSO}, z2_{\rm QSO})$, can be computed as
\begin{equation} \label{eq:5}
P_{\rm QSOz} = \frac{\int_{z1_{\rm QSO}}^{z2_{\rm QSO}} p_{\rm QSO}(z) dz}{\int p_{\rm QSO}(z) dz}\ ,
\end{equation}
which is used to quantify the uncertainty of photoz-QSO. 

Similar to the quasar class, the PDF of the galaxy class is
\begin{equation} \label{eq:6}
  p_{\rm Galaxy}(z) = \mathcal{L}_{\rm Galaxy}(z) N_{\rm Galaxy}(z).
\end{equation}
The identified photo-$z$ of galaxies is denoted as photoz-Galaxy, and the probability that the true redshift is located within $(z1_{\rm Galaxy}, z2_{\rm Galaxy})$ is
\begin{equation} \label{eq:7}
P_{\rm Galaxyz} = \frac{\int_{z1_{\rm Galaxy}}^{z2_{\rm Galaxy}} p_{\rm Galaxy}(z) dz}{\int p_{\rm Galaxy}(z) dz}
.
\end{equation}

The PDF of stars is
\begin{equation} \label{eq:8}
  p_{\rm Star}(c) = \mathcal{L}_{\rm Star}(c) N_{\rm Star}(c).
\end{equation}

The total probabilities of quasar, galaxy, and star are
\begin{eqnarray} \label{eq:9}
  p_{\rm QSO} &=& \int p_{\rm QSO}(z) dz, \nonumber \\
  p_{\rm Galaxy} &=& \int p_{\rm Galaxy}(z) dz, \\
  p_{\rm Star} &=& \int p_{\rm Star}(c) dc. \nonumber
\end{eqnarray}

Therefore, the normalized probability of an object being a quasar is expressed as
\begin{equation} \label{eq:10}
P_{\rm QSO} = \frac{p_{\rm QSO}}{p_{\rm QSO} + p_{\rm Star} + p_{\rm Galaxy}}.
\end{equation}

\begin{figure*}[!ht]
\centering
\includegraphics[width=0.8\textwidth]{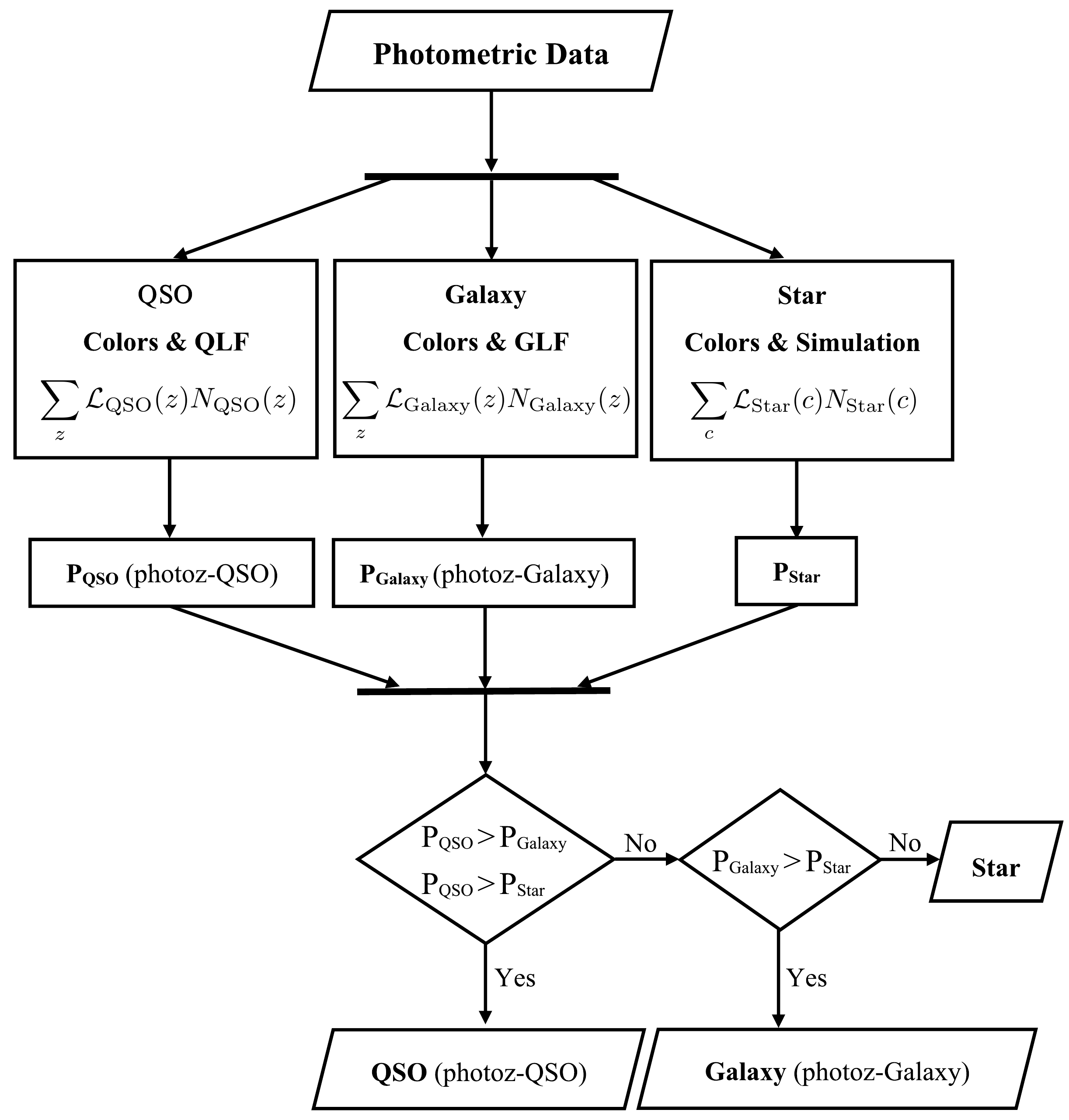}
\caption{Quasar candidate selection flowchart. For a given object with relative fluxes and magnitudes, we calculate the probability of the object being a quasar, a galaxy, or a star. We compare these probabilities, and classify the candidate into quasar, galaxy, or star. We also obtain photoz-QSO for quasar candidates and photoz-Galaxy for galaxy candidates.
\label{fig:flowchart}}
\end{figure*}

The quasar candidate selection flowchart is shown in Figure \ref{fig:flowchart}. For a given object with relative fluxes and magnitudes, we calculate the posterior probability of the object being a quasar, a galaxy, or a star combining their likelihood and prior probabilities. We compare these probabilities, and classify the candidate as a quasar, a galaxy, or a star when $P_{\rm QSO}$, $P_{\rm Galaxy}$, or $P_{\rm Star}$ is the maximum probability, respectively. By construction, these three probabilities are normalized to have a unity sum, i.e., $P_{\rm QSO} + P_{\rm Galaxy} + P_{\rm Star} = 1$. We also obtain photoz-QSO for quasar candidates and photoz-Galaxy for galaxy candidates.

\subsection{Proper Motions and Parallaxes}
It has been shown that proper motion and parallax detections from Gaia can help reduce stellar contamination in photometric quasar selection \citep[e.g., ][]{Shu_etal_2019, Calderone2019, Wolf2020}. We define the parallax significance, ${\rm PLXSIG}$, as\\
\begin{equation}
    {\rm \abs{\frac{parallax}{parallax\_error}}}\ ,
\end{equation}
where parallax\_error is the measurement uncertainty of parallax. 

Following \citet{Hambly2008}, we define the proper motion significance, ${\rm PMSIG}$, as\\
\begin{equation}
     {\rm \frac{pmra^2+pmdec^2}{\sqrt{(pmra^2~pmra\_error^2 + pmdec^2~pmdec\_error^2)}}}
\end{equation}
where pmra (pmra\_error) is the proper motion (measurement error) in right ascension, and pmdec (pmdec\_error) is the proper motion (measurement error) in declination\footnote{This PMSIG definition neglects correlated errors in pmra and pmdec.}. We use PLXSIG$<5$ and PMSIG$<5$ as additional criteria in our quasar target selection to exclude stars.

\begin{figure*}
 \subfigure{
  \includegraphics[width=3.6in]{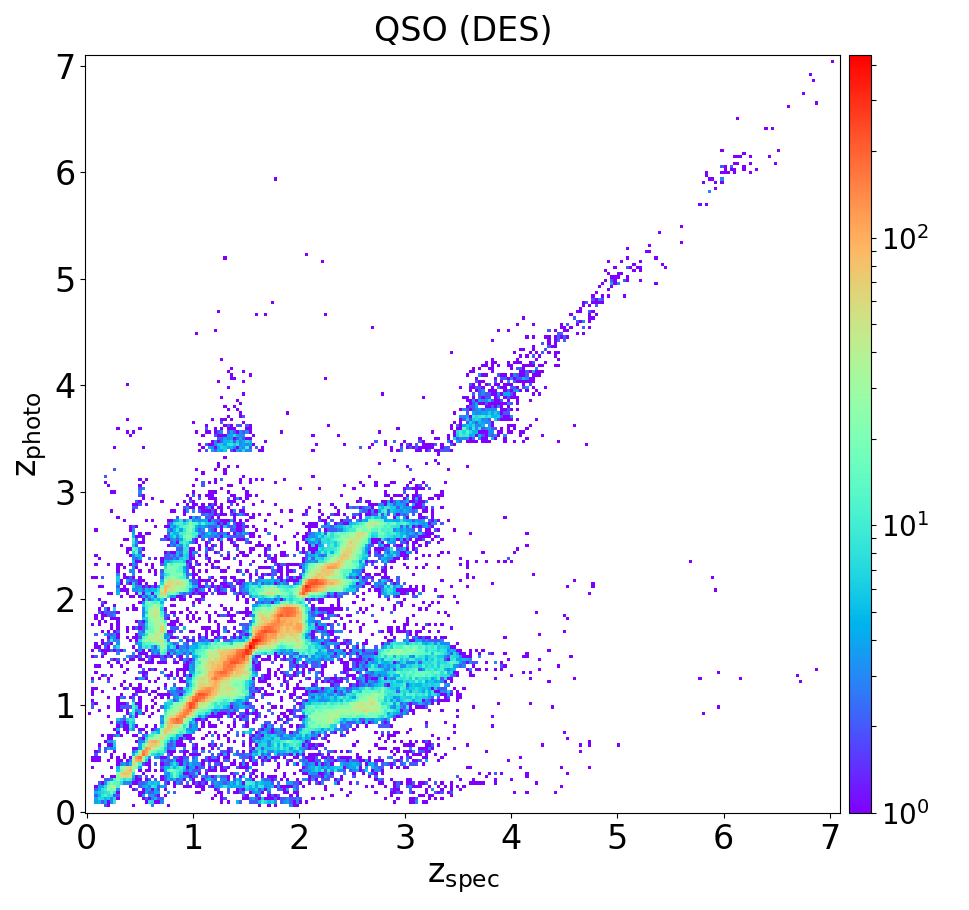}}
 \subfigure{
 \hspace{-0.5cm}
  \includegraphics[width=3.6in]{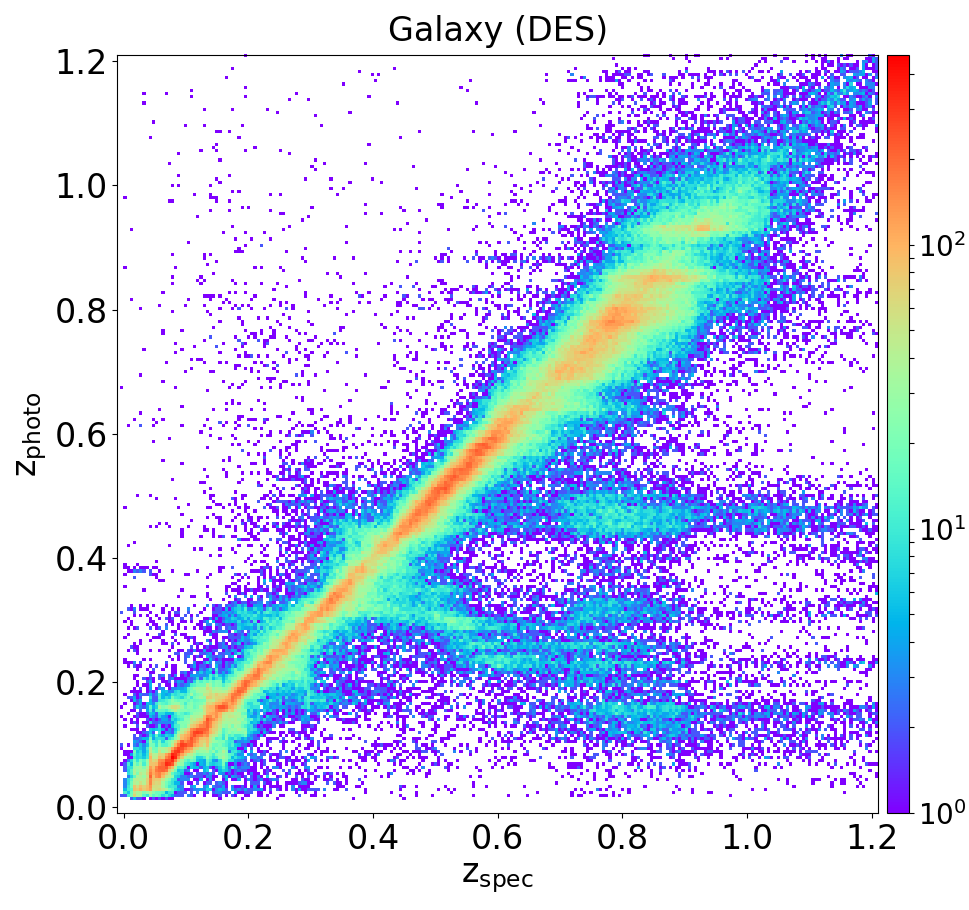}}
 \subfigure{
  \includegraphics[width=3.6in]{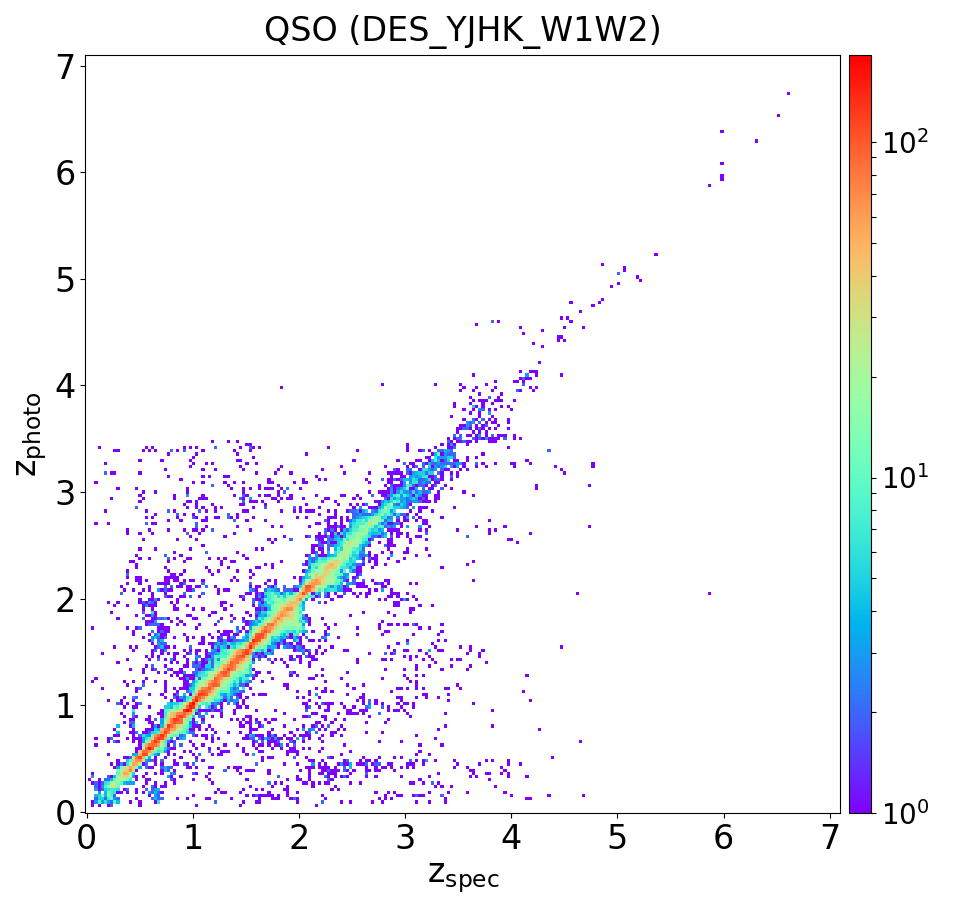}}
 \subfigure{
 \hspace{-0.5cm}
  \includegraphics[width=3.6in]{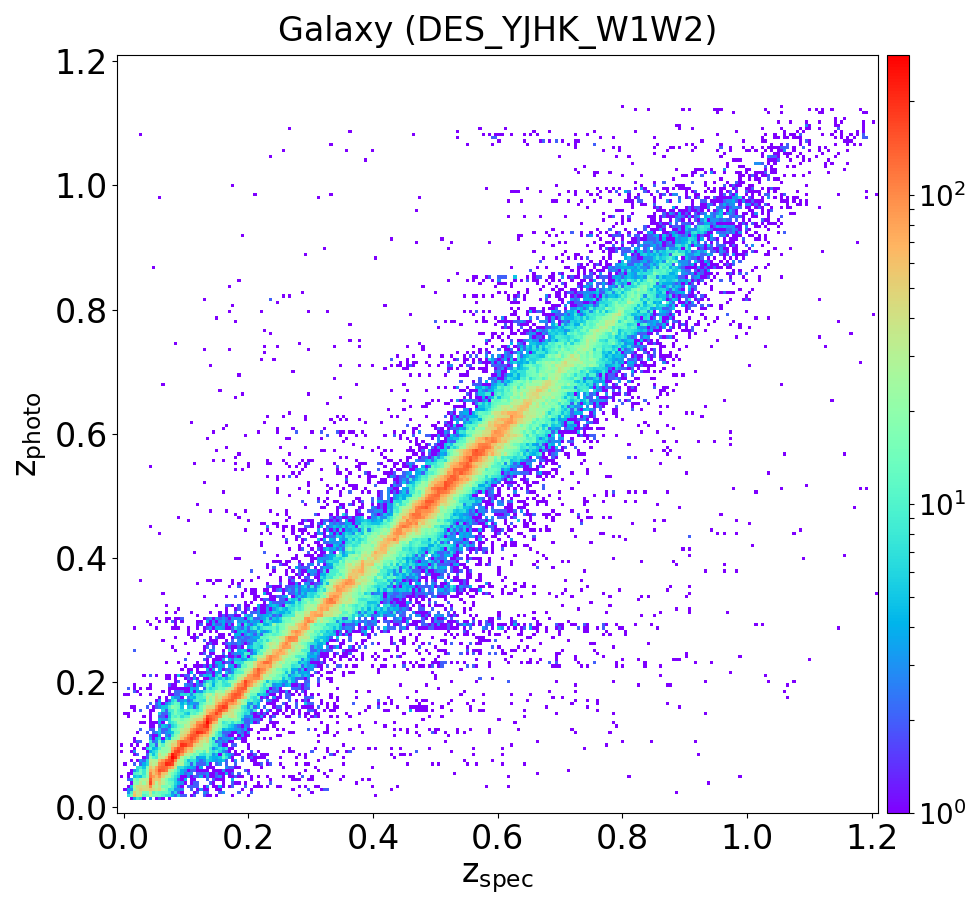}}\\
\caption{Comparison between photo-zs and the spectroscopic redshifts for quasars (left panels) and galaxies (right panels). The upper panels show the results with the fewest bands, only using 5-band DES photometry; the bottom panels show the results with the most bands, using all available photometry from optical, NIR, and MIR (i.e., the DES\_YJHK\_W1W2 combination). The colormap indicates the source number at each grid point. For quasars, there is degeneracy between $z \sim 0.8$ and $z \sim 2-3$, and this problem is alleviated by combining optical data with infrared photometry.
\label{fig:rsrp}}
\end{figure*}

\begin{figure}
\centering
\includegraphics[width=0.5\textwidth]{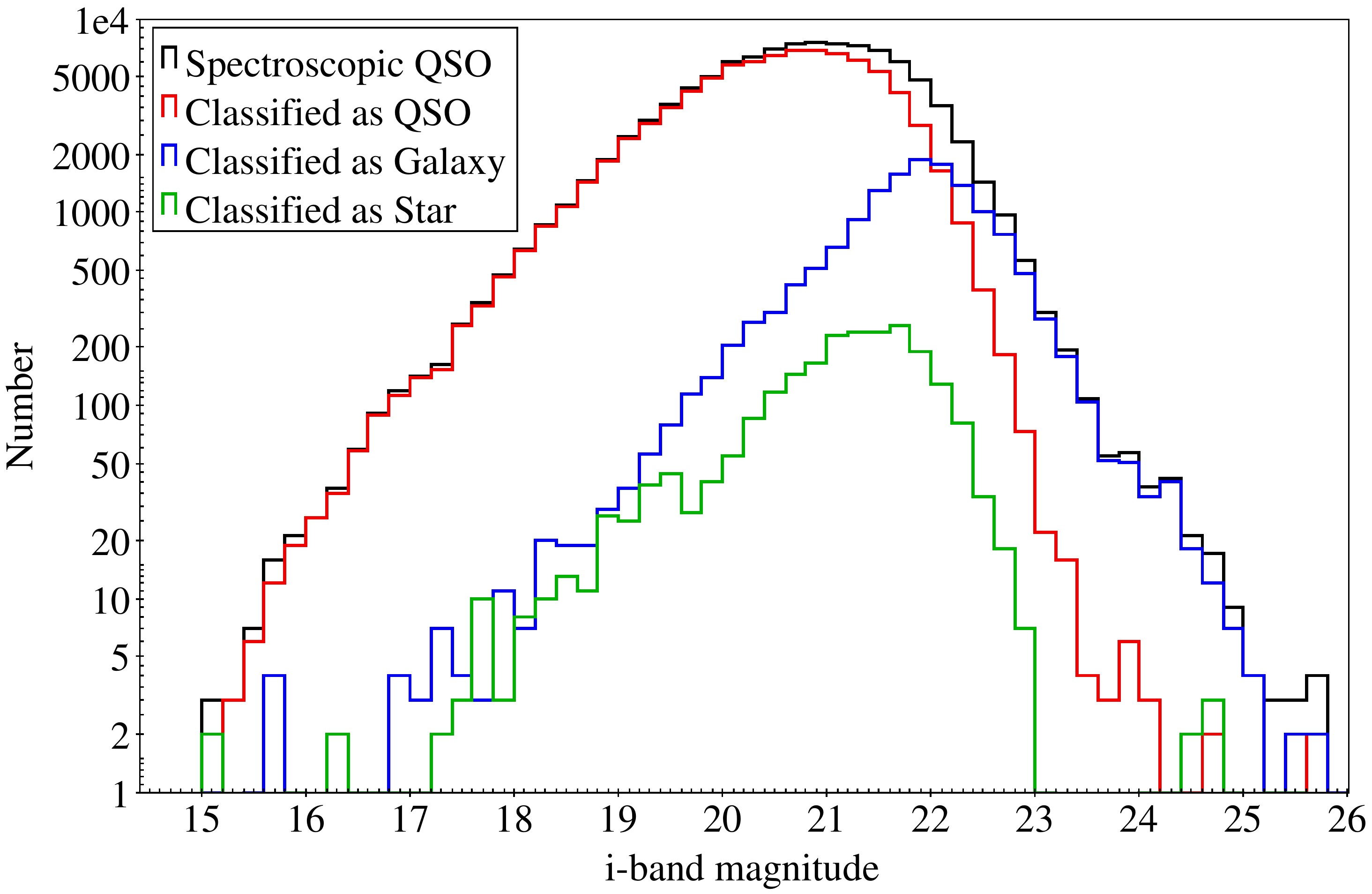}
\caption{Distribution of $i-$band magnitudes for spectroscopically known quasars (black). The red, blue, and green lines are sources classified as quasars, galaxies, and stars, respectively. The y-axis is in logarithmic scale. 
\label{fig:hist_logn}}
\end{figure}

\begin{table*}[h!]
\setlength{\tabcolsep}{1.2pt}
\caption{Regression and Classification Results for Spectroscopically Known Objects}
\begin{tabular}{l|c|c>{\bf}cc>{\bf}ccc|c>{\bf}ccc>{\bf}cc|ccc>{\bf}c}
\hline \hline
Combination & $\rm N_{\rm b}$ & \multicolumn{6}{c|}{QSO} & \multicolumn{6}{c|}{Galaxy} & \multicolumn{4}{c}{Star} \\
  &  & N & $R_{0.1}$ & $\sigma(\Delta z)$ & Q & G & S & N & $R_{0.1}$ & $\sigma(\Delta z)$ & Q & G & S & N & Q & G & S \\
\hline
DES\_YJHK\_W1W2 & 11 & 33805 & 92.2\% & 0.147  & 94.7\% & 4.9\% & 0.4\% & 134163 & 98.1\% & 0.035  & 0.5\% & 99.3\% & 0.2\% & 135975 & 0.1\% & 3.6\% & 96.3\% \\
DES\_YJHK\_W1 & 10 & 34200 & 90.9\% & 0.157  & 91.2\% & 6.7\% & 2.1\% & 139134 & 97.9\% & 0.038  & 0.5\% & 99.2\% & 0.3\% & 144598 & 0.1\% & 2.7\% & 97.1\% \\
DES\_JHK\_W1W2 & 10 & 53448 & 90.7\% & 0.157  & 92.8\% & 6.5\% & 0.7\% & 190258 & 97.6\% & 0.040  & 0.4\% & 99.2\% & 0.4\% & 176229 & 0.1\% & 4.4\% & 95.6\% \\
DES\_JK\_W1W2 & 9 & 61477 & 89.2\% & 0.160  & 91.3\% & 7.7\% & 1.0\% & 207236 & 97.1\% & 0.044  & 0.4\% & 99.2\% & 0.4\% & 180532 & 0.1\% & 3.4\% & 96.6\% \\
DES\_JHK\_W1 & 9 & 54467 & 89.2\% & 0.164  & 87.8\% & 9.6\% & 2.6\% & 199921 & 97.1\% & 0.046  & 0.5\% & 99.2\% & 0.4\% & 187735 & 0.1\% & 4.4\% & 95.5\% \\
DES\_YJHK & 9 & 34601 & 88.7\% & 0.187  & 81.9\% & 13.7\% & 4.4\% & 141975 & 96.9\% & 0.045  & 0.4\% & 99.2\% & 0.5\% & 147780 & 0.2\% & 4.4\% & 95.4\% \\
DES\_JK\_W1 & 8 & 62875 & 87.7\% & 0.170  & 85.8\% & 11.2\% & 3.0\% & 219002 & 96.3\% & 0.051  & 0.4\% & 99.2\% & 0.5\% & 193026 & 0.1\% & 3.0\% & 96.9\% \\
DES\_J\_W1W2 & 8 & 69830 & 87.2\% & 0.169  & 89.0\% & 10.4\% & 0.6\% & 214337 & 96.1\% & 0.051  & 0.3\% & 99.3\% & 0.4\% & 181408 & 0.1\% & 3.1\% & 96.8\% \\
DES\_JHK & 8 & 55462 & 86.0\% & 0.201  & 76.3\% & 17.5\% & 6.2\% & 204590 & 95.9\% & 0.053  & 0.4\% & 99.1\% & 0.5\% & 192853 & 0.1\% & 14.6\% & 85.3\% \\
DES\_W1W2 & 7 & 86947 & 85.5\% & 0.166  & 81.9\% & 17.6\% & 0.5\% & 237998 & 93.9\% & 0.067  & 0.2\% & 99.2\% & 0.6\% & 182659 & 0.1\% & 3.8\% & 96.1\% \\
DES\_J\_W1 & 7 & 72748 & 84.6\% & 0.184  & 82.4\% & 15.8\% & 1.8\% & 233336 & 94.9\% & 0.060  & 0.4\% & 99.1\% & 0.5\% & 195228 & 0.1\% & 1.6\% & 98.3\% \\
DES\_JK & 7 & 64192 & 83.9\% & 0.208  & 73.5\% & 18.2\% & 8.3\% & 224375 & 95.2\% & 0.058  & 0.3\% & 99.0\% & 0.7\% & 198569 & 0.1\% & 3.8\% & 96.1\% \\
DES\_W1 & 6 & 94277 & 82.8\% & 0.179  & 70.7\% & 27.9\% & 1.4\% & 292647 & 91.5\% & 0.080  & 0.3\% & 99.1\% & 0.6\% & 197247 & 0.2\% & 2.0\% & 97.8\% \\
DES\_J & 6 & 75397 & 77.3\% & 0.244  & 57.8\% & 23.1\% & 19.1\% & 241349 & 93.6\% & 0.068  & 0.2\% & 98.6\% & 1.2\% & 208799 & 0.2\% & 3.3\% & 96.5\% \\
DES & 5 & 102321 & 72.2\% & 0.273  & 40.6\% & 42.3\% & 17.1\% & 318138 & 90.0\% & 0.085  & 0.2\% & 98.2\% & 1.6\% & 219857 & 0.3\% & 7.2\% & 92.5\% \\
\hline
\hline
\end{tabular}
\label{table:results}
\tablecomments{$\rm N_{\rm b}$ is the number of bands. N is the number of objects. $R_{0.1}$ is the fraction of objects with $|\Delta z|$ smaller than 0.1, where $\Delta z = (z_s - z_p)/(1+z_s)$. $\sigma(\Delta z)$ is the standard deviation of $\Delta z$. The three columns ``Q'', ``G", and ``S" stands for fractions of objects classified as quasars, galaxies, and stars.
 The table is ranked by the $R_{0.1}$ value of QSO (from high to low). For visual aid, we use bold fonts to highlight the most useful columns (larger values better).}
\end{table*}

\begin{table*}
\caption{QSO Selection Criteria \label{tab:criteria}}
  \centering
    \begin{tabular}{ll|rr|rr}
    \hline
    \hline
    \multicolumn{1}{l}{Criteria} & \multicolumn{1}{c}{} & \multicolumn{2}{|c|}{Candidates} & \multicolumn{2}{|c}{Known QSOs in DES-DR2}\\ 
    & & \multicolumn{1}{c}{\multirow{2}{*}{All}} & $P_{\rm QSO} > P_{\rm Star}$ \&  & \multicolumn{1}{c}{\multirow{2}{*}{All}} & \multicolumn{1}{c}{\multirow{2}{*}{$i<21.5$}} \\ 
    &  &     & $P_{\rm QSO} > P_{\rm Galaxy}$   &     &          \\
    \hline
%----------------------------
All & & 691,483,608 \ (100\%) & 1,471,001 \ (100\%) & 102,321 \ (100\%) & 84,280 \ (100\%) \\
\hline
%----------------------------
% The max PSF SNR in optical bands is higher than 5
$\rm SN\_MAX\_PSF > 5$ & (1) & 391,279,183 (56.6\%) & 1,402,955 (95.4\%) & 102,173 (99.9\%) & 84,174 (99.9\%) \\
%----------------------------
% The SNR is higher than 3 in at least 3 bands \\
$\rm SN3 > 2$ & (2) & 375,038,065 (54.2\%) & 1,375,613 (93.5\%) & 102,162 (99.8\%) & 84,174 (99.9\%) \\
%----------------------------
% The $\rm IMAFLAGS\_ISO$ is zero in all DES bands\\
$\rm IMAFLAGS\_ISO = 0$ & (3) &  372,958,657 (53.9\%) & 1,358,057 (92.3\%) & 101,890 (99.6\%) & 83,966 (99.6\%) \\
%----------------------------
PLXSIG$<$5 \& PMSIG$<$5 & (4) & 358,069,257 (51.8\%) & 1,352,947 (92.2\%) & 101,706 (99.4\%) & 83,782 (99.4\%) \\
%----------------------------
$P_{\rm QSO} > P_{\rm Star}$ \& $P_{\rm QSO} > P_{\rm Galaxy}$ & (5) & 1,352,947 (0.196\%) & 1,352,947 (92.2\%) & 84,978 (83.1\%) & 76,424 (90.7\%) \\ 
%----------------------------
\hline
%----------------------------
$P_{\rm QSO} > 0.7$ & (6) & 945,860 (0.137\%) & 945,860 (64.3\%) & 79,811 (78.0\%) & 73,802 (87.6\%) \\
%----------------------------
\hline
\end{tabular}
\tablecomments{We use criteria (1)-(5) for our fiducial quasar catalog, which includes 1,352,947 quasar candidates (see \S\ref{sec:selection} for details on the criteria). Since $P_{\rm QSO} + P_{\rm Star} + P_{\rm Galaxy} = 1$, criterion (5) implies $P_{\rm QSO} > 1/3$. 
For a high-completeness selection, we recommend to use our fiducial quasar catalog, selected with criteria (1)-(5). For a higher efficiency (purity) selection while maintaining a high completeness ($>85\%$ at $i<21.5$), we recommend to add criterion (6) of ${P_{\rm QSO}} > 0.7$. Each selection step also includes all previous criteria. The percentages in parentheses are the fractions of objects among the full sample.
}
\end{table*}

\section{Results} \label{sec:results}

\subsection{Photo-$z$ Regression and Classification Results}
Table \ref{table:results} summarizes the photo-$z$ regression and classification results for spectroscopically confirmed objects (quasars, galaxies, and stars) for different photometric band combinations. In total, we used 15 most frequent photometric band combinations. In general, the photo-$z$ regression and classification results are better when more bands are used, as expected.

The difference between the photo-$z$ ($z_p$) and the spectroscopic redshift ($z_s$) is quantified as $|\Delta z| \equiv |z_s - z_p| / (1 + z_s)$. The photo-$z$ accuracy $R_{0.1}$ is the fraction of objects in a test sample with $|\Delta z|$ smaller than 0.1. A larger $R_{0.1}$ represents a higher photo-$z$ accuracy. In addition, a smaller standard deviation of $\Delta z$ measured for the test sample, $\sigma(\Delta z)$, would indicate that the photo-$z$ result is better overall. When using $grizY$ bands from DES, combined with all available IR bands ($YJHK$ in NIR and $W1W2$ in MIR), the photo-$z$ accuracy is as good as 92.2\% for quasars and 98.1\% for galaxies for our spectroscopic training samples. The standard deviation of $\Delta z$, $\sigma(\Delta z)$, is 0.147 and 0.035 for quasars and galaxies, respectively. 

As shown in Table \ref{table:results}, with fewer bands, $R_{0.1}$ decreases and $\sigma(\Delta z)$ increases for both quasars and galaxies, as expected. When only using DES bands, $R_{0.1}=72.2\%$ for quasars and 90.0\% for galaxies; $\sigma(\Delta z)= 0.273$ and 0.085 for quasars and galaxies, respectively. For comparison, $R_{0.1}=74.2\%$ and $\sigma(\Delta z)= 0.27$ for quasars when only using SDSS $ugriz$ bands \citep{Yang2017}. The photoz-QSO accuracy using DES photometry is slightly worse than that using SDSS photometry because there is no $u$ band in DES, which is useful for quasar photo-$z$ calculation at low redshift. As shown in \citet{Yang2017}, $R_{0.1}$ is 72.8\% and $\sigma(\Delta z)$ is 0.31 using the XDQSOz method \citep{Bovy2012}. Using our algorithm and the DES photometric data, the photo-$z$ accuracy is comparable or slightly better than the XDQSOz algorithm using SDSS photometry. 

Figure \ref{fig:rsrp} shows the photo-$z$ versus the spectroscopic redshift with the fewest bands (DES only, upper panels) and most bands (DES\_YJHK\_W1W2, bottom panels) for quasars (left panels) and galaxies (right panels). The color-map shows the number density. For quasars, using DES data alone, there is an apparent degeneracy between $z\sim2.2$ and $z\sim0.8$, since the DES colors of quasars at both redshifts are similar as the C III] (Mg II) line shifts into the $g$ band at $z\sim2.2$ ($z\sim0.8$). This degeneracy is resolved with the inclusion of NIR and MIR data. For galaxies, there is some degeneracy at $z>0.5$ and this problem is also alleviated by including NIR and MIR data.

Our algorithm not only calculates quasar and galaxy photo-$z$, but also classifies quasars, galaxies, and stars based on the maximum probability. In Table \ref{table:results}, we show the fraction of objects classified as quasars, galaxies and stars in the spectroscopic training samples. We used the $P_{\rm QSO}$, $P_{\rm Galaxy}$, and $P_{\rm Star}$ parameters for the classification. As illustrated in Figure \ref{fig:flowchart}, a target is classified as a quasar when its $P_{\rm QSO}$ is higher than $P_{\rm Galaxy}$ and $P_{\rm Star}$ (thus the normalized probability $P_{\rm {QSO}}$ in Equation \ref{eq:10} is larger than $1/3$). We successfully classify 94.7\% of quasars, 99.3\% of galaxies, and 96.3\% of stars when all bands are available. Fewer quasars are mis-classified as stars when including MIR photometry since stars usually radiate thermal emission and are faint in the MIR. At the faint end, without NIR and/or MIR photometry, more quasars are mis-classified as galaxies due to contamination from their host galaxies. 

Figure \ref{fig:hist_logn} shows the distribution of $i$-band magnitude for the 102321 spectroscopically confirmed quasars in DES footprint along with our photometric classifications. In this figure, we use all available bands, and 83\%, 14\%, and 2\% of them are classified as quasars, galaxies, and stars, respectively. Quasars at the faint end, especially at $i>22$, might be misclassified as galaxies.

Stars and galaxies mis-classified as quasars will decrease the purity of the selected quasar candidate sample. Using our benchmark sample of spectroscopically confirmed galaxies and stars, only a small fraction (0.1\%-0.3\%) of stars are mis-classified as quasars, and a small fraction (0.2\%-0.5\%) of galaxies are mis-classified as quasars (see Table \ref{table:results}). These contamination rates are based on the most loose quasar selection criteria. Using a higher $P_{\rm QSO}$ cut, the contamination from stars and galaxies can be further reduced. Of course, the absolute contamination fraction depends on the densities of stars and galaxies in the targeting field. In \S\ref{sec:selection}, we show the full selection criteria, as well as the completeness and efficiency (purity) for our quasar selection for typical high Galactic-latitude fields. 

\begin{figure*}%[htbp]
 \subfigure{
  \includegraphics[width=3.6in]{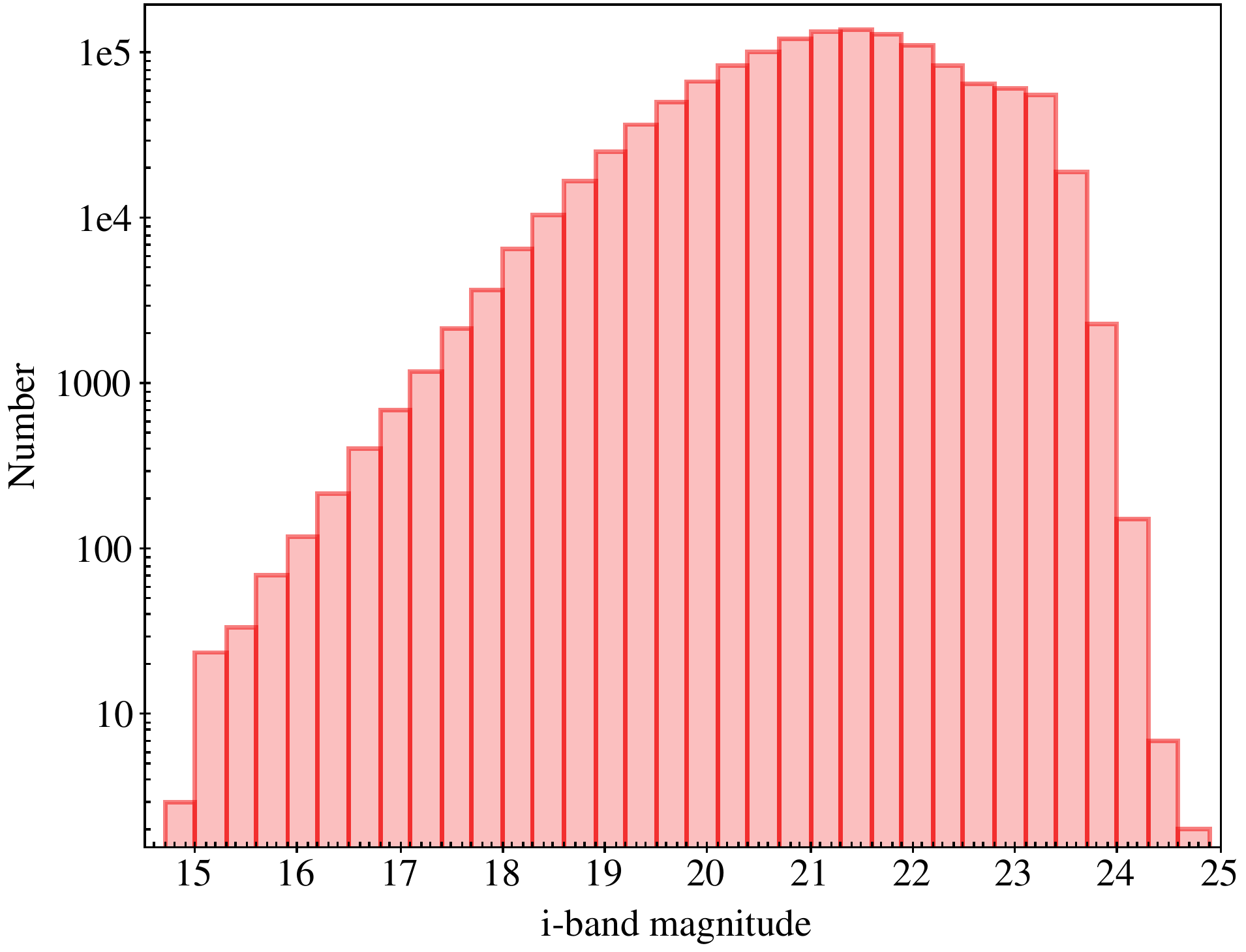}}
 \subfigure{
  \includegraphics[width=3.6in]{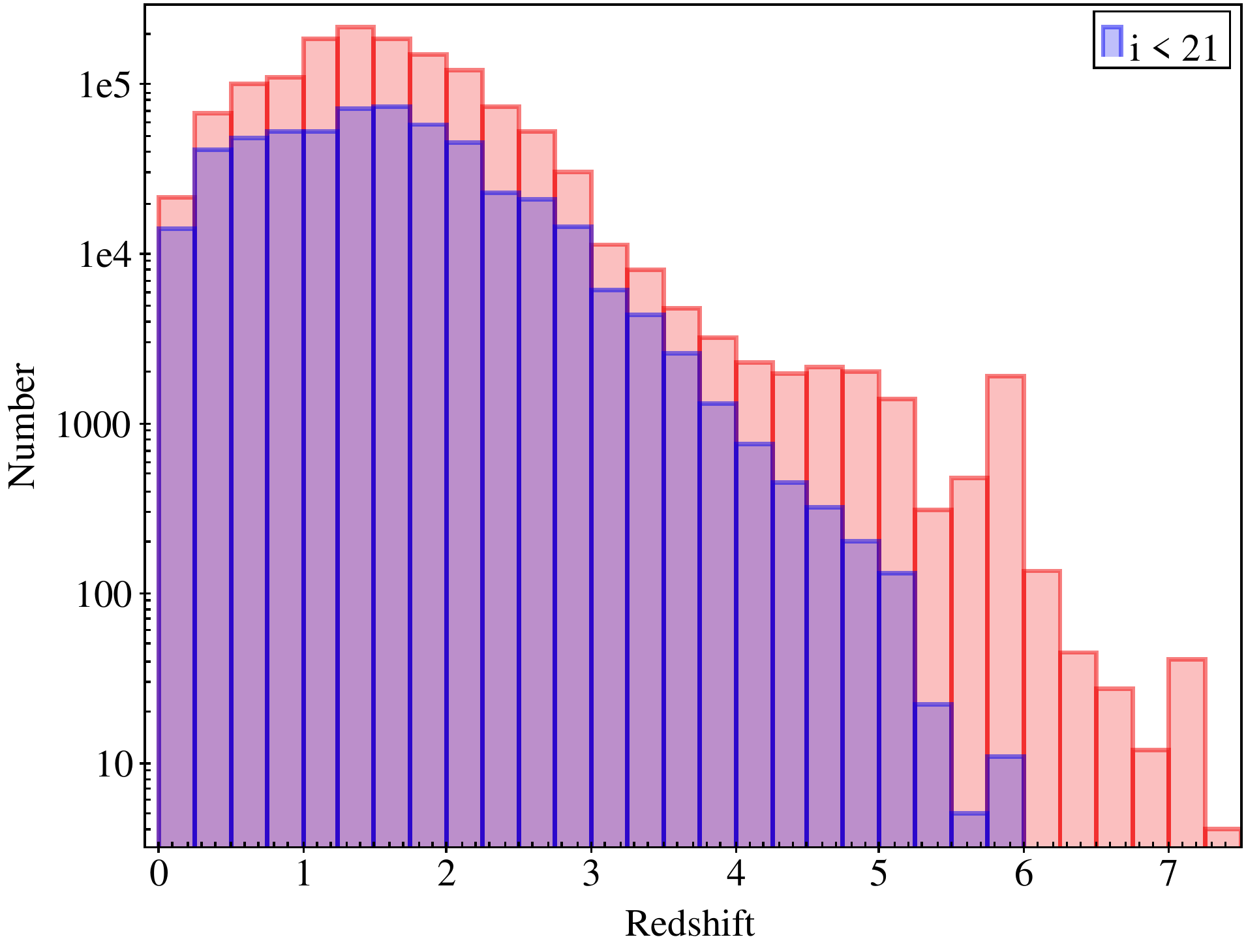}}\\
\caption{Distributions of $i$-band magnitudes (left panel) and photometric redshifts (right panel) of the 1,352,947 quasar candidates in the DES wide footprint. The y axes are in logarithmic scales. The blue histogram in the right panel are quasar candidates with $i < 21$.
\label{fig:hist_candi}}
\end{figure*}

\subsection{Quasar Candidates} \label{sec:selection}
We now perform quasar target selection over the 5000 deg$^2$ DES wide-field area. Table \ref{tab:criteria} summarizes the steps to select quasar candidates. We use the following criteria to optimize the quasar selection:

\begin{enumerate}
  \item The maximum S/N in five DES bands is greater than 5, ${\rm SN\_MAX\_PSF > 5}$. 
  
  \item At least two DES bands have a S/N greater than 3, ${\rm SN3>2 }$.
  
  \item We request a baseline quality criterion of IMAFLAGS\_ISO = 0 in all DES bands.
  
  \item The Gaia proper motion significance, ${\rm PMSIG}$, and parallax significance, ${\rm PLXSIG}$, are smaller than 5.
  \item The Skewt-QSO probability of quasars, ${P_{\rm QSO}}$, is larger than those of stars, ${P_{\rm Star}}$, and galaxies, ${P_{\rm Galaxy}}$, i.e., ${P_{\rm QSO} > P_{\rm Star} }$ and ${P_{\rm QSO} > P_{\rm Galaxy} }$; thus ${P_{\rm QSO}} > 1/3$ by construction.
\end{enumerate}

\begin{figure}
\centering
\includegraphics[width=0.5\textwidth]{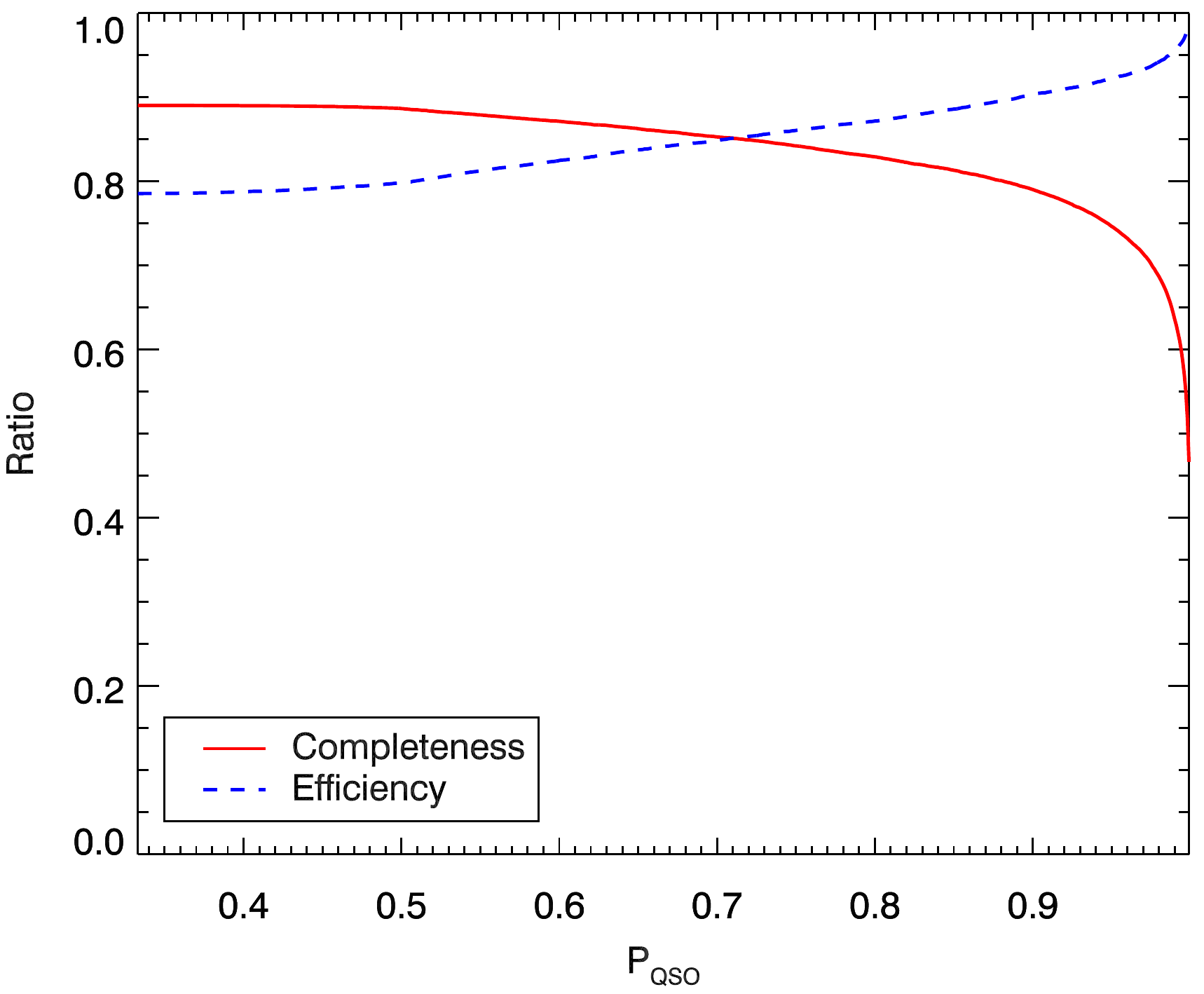}
\caption{Completeness (red solid line) and efficiency (blue dashed line) as a function of $P_{\rm QSO}$ for spectroscopically confirmed quasars in the S82 region ($r<21.5$).
\label{fig:com_eff}}
\end{figure}

\begin{figure}
\centering
\includegraphics[width=0.48\textwidth]{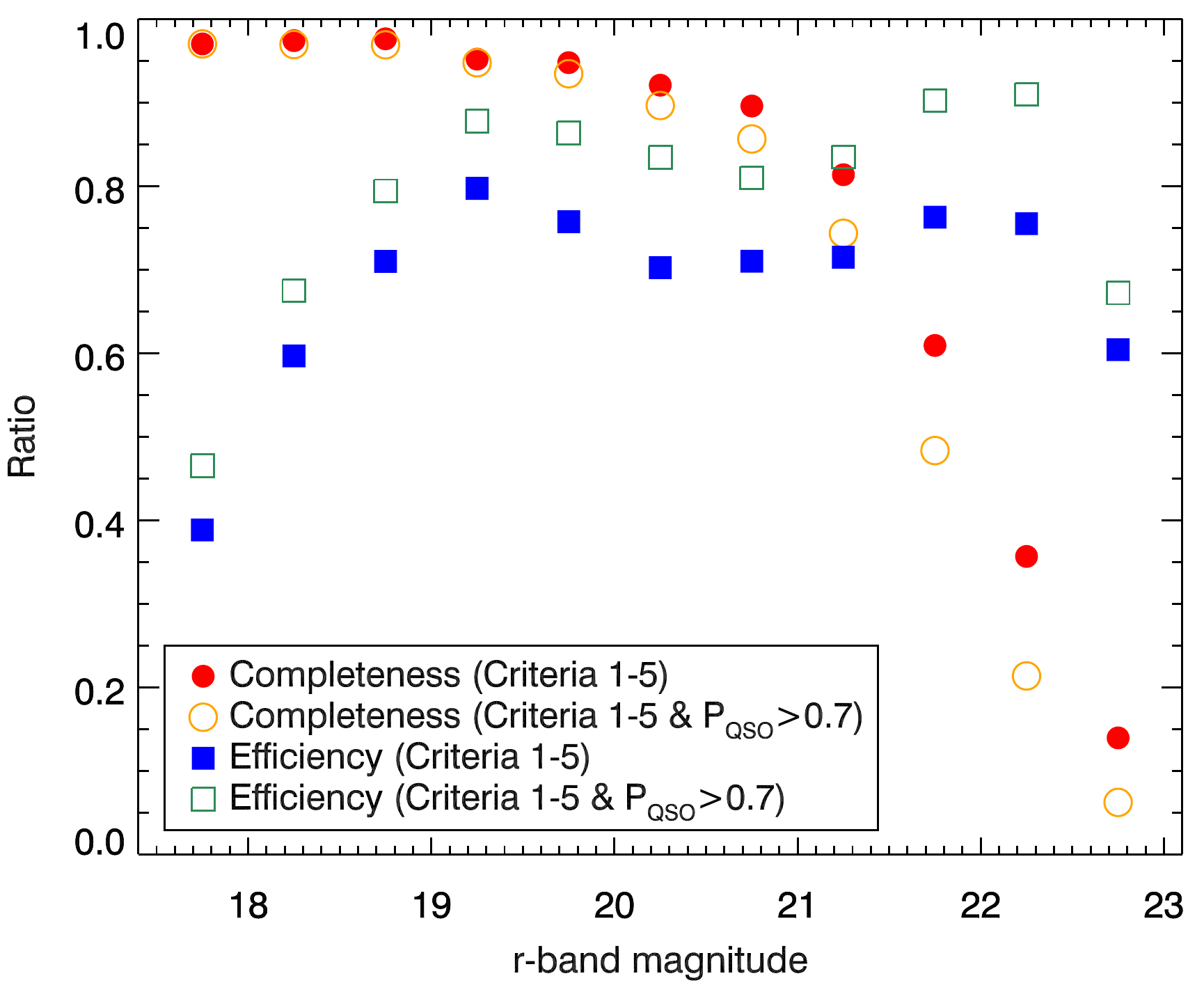}
\caption{Completeness and efficiency as function of $r$-band magnitude, using spectroscopically confirmed quasars in the S82 region. The solid red circles and blue squares show results with selection criteria (1)-(5), and the open orange circles and green squares are results with one more criterion $P_{\rm QSO}>0.7$. }
\label{fig:com_eff_mag}
\end{figure}

\begin{figure*}
\centering
\includegraphics[width=0.9\textwidth]{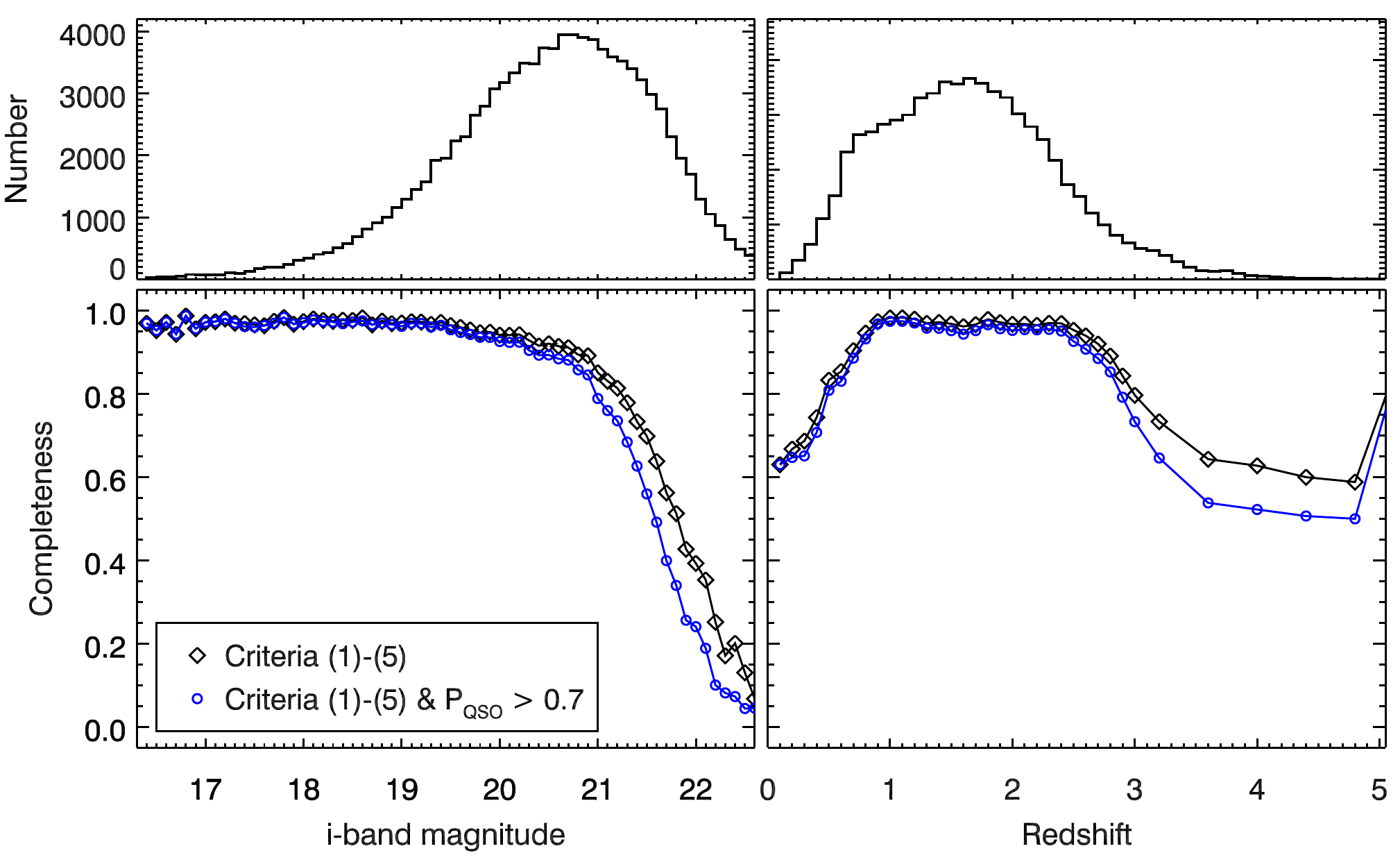}
\caption{Completeness as function of $i$-band magnitude (left panels) and redshift (right panels) for spectroscopically confirmed quasars. The top panels show the number distribution. The black diamonds/lines show the results from criteria (1)--(5). The blue circles/lines represent results from criteria (1)--(6).
\label{fig:mag_z}}
\end{figure*}

In total, there are 691,483,608 sources in the DES DR2 coadded photometric catalog. Among these sources, there are 1.47, 645.88, and 44.13 million sources classified as quasars, galaxies, and stars, respectively, using the Skewt-QSO probabilities only (i.e., criterion 5). Using criteria (1)--(5) above, we photometrically classify 1,352,947 as quasar candidates, 334,484,173 as galaxy candidates and 36,950,258 as star candidates (criterion 4 were not applied to star candidates). Figure \ref{fig:hist_candi} shows the $i$-band magnitude and redshift distributions of the 1.35 million quasar candidates. In both panels, the y axes are in logarithmic scale. The left panel shows that the targets are highly complete at $i<21$ with log-linear number increasing from bright to faint magnitude. In the right panel, the redshift distribution peaks around 1.5. The QLF studies show that the number density of luminous quasars peaks between redshifts 2 and 3 \citep[e.g., ][]{Richards2006}. For quasars with the same absolute magnitude, the apparent magnitude becomes fainter as redshift increases, and in the fainter regime the selection completeness decreases, so the redshift peak moves towards lower redshift. The blue histogram in the right panel shows the quasar candidates with $i<21$. With higher completeness at $i<21$, the redshift distribution between 0.5 and 2.2 becomes flatter and the peak moves slightly to higher redshift. There are few $z>6$ quasar candidates with $i>21$ as the Lyman-$\alpha$ emission line drops out of $i$ band at $z>6$.

Among the set of selection criteria, the first criterion excludes 43.4\% of DES sources. The second, third, and forth criteria further exclude 4.8\% of DES sources. The most crucial criterion is the fifth criterion from the Skewt-QSO probability, excluding 51.6\% of DES sources. Quasars are normally point-like sources, but low-redshift quasars and faint quasars can be extended sources. Therefore, we did not perform any morphological cuts based on DES imaging.

For higher selection efficiency (purity), we can adopt higher ${P_{\rm QSO}}$ thresholds. We tested the completeness and efficiency (purity) of quasar selection in the Stripe 82 (S82) region of SDSS, where the spectroscopic completeness of photometric objects is relatively high. Specifically, we use the S82 region with ${\rm R.A.} < 45^{\circ}$ or ${\rm R.A.} > 317^{\circ}$ and $|{\rm Decl.}|<1.25^{\circ}$. Since the completeness and efficiency vary with magnitude and decrease dramatically at the faint end, here we use quasars brighter than $r=21.5$, which is appropriate for current spectroscopic quasar surveys. Following \citet{Yang2017}, the efficiency (purity) is calculated based on the quasar number estimated from the QLF as \\
\begin{equation} \label{eq:eff}
  {\rm efficiency} = \frac{N_{\rm QLF}(r) \times {\rm completeness}(r)}{N_{\rm candidates}(r)}\ ,
\end{equation}
where $N_{\rm QLF}(r)$ is the number of quasars per deg$^2$ calculated from the QLF \citep{NPD2016}.
Figure \ref{fig:com_eff} shows the completeness and efficiency for different values of ${P_{\rm QSO}}$ using the S82 spectroscopically confirmed quasar sample. 
As we adopt a higher ${P_{\rm QSO}}$ threshold, the completeness decreases and the efficiency of the selection increases. With our fiducial criteria (1-5), there are very few spectroscopically confirmed quasars (0.4\%) and quasar candidates (2\%) with $1/3<P_{\rm QSO} \leq 0.5$. The completeness and efficiency are both high ($\sim$85\%) when using a ${P_{\rm QSO}}$ threshold of 0.7.
Therefore, for a high-completeness selection, we recommend to use our fiducial quasar catalog, selected using criteria (1)-(5). For a higher efficiency selection while maintaining a high completeness ($\sim85\%$), we recommend to add one more criterion of ${P_{\rm QSO}} > 0.7$, which results in 0.95 million quasar candidates.

Figure \ref{fig:com_eff_mag} shows the completeness and efficiency as function of $r$-band magnitude. The completeness using one more criterion of $P_{\rm QSO}>0.7$ is lower  than that only using criteria (1)-(5), while the efficiency behaves in the opposite sense. The completeness falls below 80\% at $r>21.5$. The drop of efficiency (purity) at the bright end is mainly due to enhanced contamination from mis-classified low-redshift bright galaxies. Our algorithm can select some weak-line or narrow-line AGNs. For example, among 5741 narrow-line AGNs in the Million Quasars Catalog (type=“K” or “N”, i.e., narrow-line quasars or narrow-line Seyferts) within the DES footprint, our algorithm selects 492 of them. Therefore, the completeness will increase and the contamination rate will decrease if we include narrow-line objects in our quasar selection. On the other hand, the measurements of the QLF are generally difficult at the bright end given the rapid decrease in the spatial density of quasars toward high luminosities. Therefore our estimated efficiency at the bright end is highly impacted by the quality of the QLF measurement.

Of course, the efficiency of quasar selection also depends on the field stellar density. In sky regions with high stellar densities, the purity would decrease, as more stars would be mis-classified as quasars (even if the fraction of stars mis-classified as quasars is as low as 0.1\%).

We also tested applying the most crucial criterion from the Skewt-QSO probability first, resulting in 1.47 million quasar candidates (2.14\% of all the source in DES DR2). The other criteria further rule out 0.017\% (118,054) sources, demonstrating that ${P_{\rm QSO}}$ is the most useful parameter to rule out contamination. The Gaia astrometry criteria rule out 5110 additional sources. In the bright regime where the Gaia detects proper motion, 4375 out of 536956 sources are rejected at $i<21$. This confirms that our Skewt-QSO probability criterion selects very few stellar contaminants with large parallaxes/proper motions. Of course, our photometric quasar sample may still contain many faint stars without reliable Gaia DR2 astrometry. 

Using probability distributions of parallax/proper motion as a prior probability or machine learning approaches as in \citet{Shu_etal_2019}  will make better use of Gaia astrometric information. However, as shown in Table \ref{tab:criteria}, the Skewt-QSO color selection has already ruled out the majority of stars, and using the additional parallax/proper motion cuts of PMSIG $<$ 5 and PLXSIG $<$ 5 only rules out $<$0.1\% additional sources of the 1.35 million quasar candidates after the Skewt-QSO criteria. In that sense, more refined parallax/proper motion cuts are unnecessary, since the primary selection of our quasar candidates is the Skewt-QSO color selection.

In Table \ref{tab:criteria}, we also list the numbers of spectroscopically confirmed quasars in the DES-DR2 source catalog that pass our selection criteria. Criteria (1)-(5) recover 83.1\% (90.7\%) of all ($i<21.5$) spectroscopically confirmed quasars. Using ${P_{\rm QSO}} > 0.7$, the completeness is 78.0\% (87.6\%) for all ($i<21.5$) quasars. 

We provide probabilities of quasars, galaxies, and stars for the entire DES DR2 coadded photometric catalog, which contains a total of 691,483,608 sources. The format of our final catalogs is described in Table~\ref{tab:catalog} for the $\sim 1.4$\,million quasar candidates and in Table~\ref{tab:catalog_all} for the full DES-DR2 source catalog. These catalogs can be downloaded from URL\footnote{\url{http://quasar.astro.illinois.edu/paper_data/DES_QSO/}}. 

\begin{deluxetable*}{llll}
\tablecaption{FITS Table Format for DES DR2 Skewt-QSO Catalog  \label{tab:catalog}}
\tablewidth{1pt}
\tablehead{
\colhead{Column Name} &
\colhead{Format} &
\colhead{Units} &
\colhead{Description}
}
\startdata
% ------------ DES information ------------
COADD\_OBJECT\_ID	& LONG64 & & Unique identifier for the coadded objects \\       
ALPHAWIN\_J2000 & DOUBLE & degree & DES Right Ascension (J2000) \\
DELTAWIN\_J2000 & DOUBLE & degree & DES Declination (J2000) \\
EXTENDED\_COADD & INT & & DES morphological object classification variable \\
 & & & $~~~~$0: high confidence pointed-like; 1: likely pointed-like; \\
 & & & $~~~~$2: likely extended; 3: high confidence extended\\
SN\_MAX\_PSF & FLOAT & & MAX SNR of the PSF mag in DES \\
SN3 & INT & & The number of bands in DES with SNR higher than 3 \\
Photometry & STRING & & DES photometry fitting to quasar and star models, PSF or AUTO \\
Band\_DES & STRING & & DES bands \\
MAG\_PSF\_GRIZY & FLOAT & mag & DES PSF magnitudes in $grizY$ bands\\
MAGERR\_PSF\_GRIZY & FLOAT & mag & DES PSF magnitude uncertainties in $grizY$ bands\\
MAG\_AUTO\_GRIZY & FLOAT & mag & DES AUTO magnitude in $grizY$ bands \\
MAGERR\_AUTO\_GRIZY & FLOAT & mag & DES AUTO magnitude uncertainties in $grizY$ bands \\
IMAFLAGS\_ISO\_GRIZY & INT & & DES flag in $grizY$ bands \\
% ------------ Gaia proper motion ------------
PLXSIG & FLOAT & & Gaia DR2 parallax significance\\
PMSIG & FLOAT & & Gaia DR2 proper motion significance\\
Separation\_Gaia & FLOAT & arcsec & Angular distance between DES and Gaia coordinates \\
CNT9 & INT & & Number of sources with a 9 arcsec radius circular aperture \\
DIST & FLOAT & arcsec & Angular distance to the closest neighbour within 9 arcsec\\
% ------------ IR ------------
Survey\_NIR & STRING & & NIR survey \\
Band\_NIR & STRING & & NIR bands \\
Nband\_NIR & INT & & Number of NIR bands \\
Separation\_NIR & FLOAT & arcsec & Angular distance between DES and NIR coordinates  \\
Mag\_YJHK & FLOAT & mag & NIR magnitudes in $YJHK$ bands (AB magnitude) \\
Magerr\_YJHK & FLOAT & mag & NIR magnitude uncertainties in $YJHK$ bands \\
Band\_WISE & STRING & & WISE bands (only use $W1$ and $W2$ bands)\\
Nband\_WISE & INT & & Number of WISE bands \\
Separation\_WISE & FLOAT & arcsec & Angular distance between DES and unWISE coordinates  \\
Mag\_W1W2& FLOAT & mag & WISE magnitudes in $W1$ and $W2$ bands (AB magnitude) \\
Magerr\_W1W2& FLOAT & mag & WISE magnitude uncertainties in $W1$ and $W2$ bands \\
% ------------ Skewt-QSO ------------
Combination & STRING & & DES, NIR, and MIR band combination \\
Reference\_Band & STRING & & DES reference band \\
$P\_{\rm QSO}$ & FLOAT & & $P_{\rm QSO}$, Skewt-QSO probability fitting to QSO models, described in Eq. \ref{eq:10} \\
photoz\_QSO & FLOAT & & Quasar photo-$z$\\
$z1\_{\rm QSO}$ & FLOAT & & The lower limit of quasar photo-$z$ \\
$z2\_{\rm QSO}$ & FLOAT & & The upper limit of quasar photo-$z$ \\
$P\_{{\rm QSO}}\_z$ & DOUBLE & & The probability of quasar photo-$z$ locating within ($z1_{\rm QSO}$, $z2_{\rm QSO}$), described in Eq. \ref{eq:5}  \\
$P\_{\rm Galaxy}$ & FLOAT & & $P_{\rm Galaxy}$, Skewt-QSO probability fitting to Galaxy models \\
photoz\_Galaxy & FLOAT & & Galaxy photo-$z$\\
$z1\_{\rm Galaxy}$ & FLOAT & & The lower limit of galaxy photo-$z$ \\
$z2\_{\rm Galaxy}$ & FLOAT & & The upper limit of galaxy photo-$z$ \\
$P\_{{\rm Galaxy}\_z}$ & DOUBLE & & The probability of galaxy photo-$z$ locating within ($z1_{\rm Galaxy}$, $z2_{\rm Galaxy}$), described in Eq. \ref{eq:7}  \\
$P\_{\rm Star}$ & FLOAT & & $P_{\rm Star}$, Skewt-QSO probability fitting to Star models  \\
$z$\_spec & DOUBLE & & {Spectroscopic redshift, if available} \\
$z$\_spec\_cat & STRING & & {Spectroscopic redshift catalog, i.e., SDSS, Milliquas, or LAMOST}\\
Class\_spec & STRING & & {Spectroscopic classification}
\enddata
\end{deluxetable*}

\section{Discussion}\label{sec:disc}

\subsection{2D completeness in the magnitude-redshift space}

Using spectroscopically confirmed quasars, we further quantify our quasar selection completeness as a function of both magnitude and redshift. As shown in \citet{Yang2017}, the photometric redshift accuracy and the classification success rate of the Skewt-QSO algorithm are high even when using different training and testing samples. 

Figure \ref{fig:mag_z} shows the completeness as a function of $i$-band magnitude (left panels) and redshift (right panels) using spectroscopically confirmed quasars. The black diamonds/lines represent the selection results using our fiducial criteria (1)-(5). The blue circles/lines represent the results using one more criterion of $P_{\rm QSO}>0.7$. The completeness is higher than 80\% at $i < 21$ for both selections. The completeness decreases rapidly at $i>21$, which is the consequence of decreasing photometric accuracy and the lack of infrared detection at the faint end. The right panel in Figure \ref{fig:mag_z} shows the completeness as a function of redshift for $i<21$ quasars. The overall completeness is $>80\%$ for quasars over $0.5<z<3$. At the low redshift end ($z<1$), the selection completeness decreases as redshift decreases, which is due to enhanced contamination from bright host galaxies of quasars at low redshift that causes the quasar not to be selected based on color. At high redshift ($z>2.5$), the completeness decreases with redshift. At $z>3$ the completeness estimation suffers from the small spectroscopic sample size, so we use a larger redshift bin of 0.4 at $z>3$, comparing to a bin of 0.1 at $z<3$, to avoid large fluctuations. In addition, these completeness estimates are based on spectroscopically confirmed quasars, thus suffering from their own selection effects and incompleteness. So the total completeness might be even lower than our estimation from the spectroscopic sample, especially at the faint end where the original spectroscopic sample suffers the most from incompleteness in selection.

\begin{figure}
\centering
\includegraphics[width=0.5\textwidth]{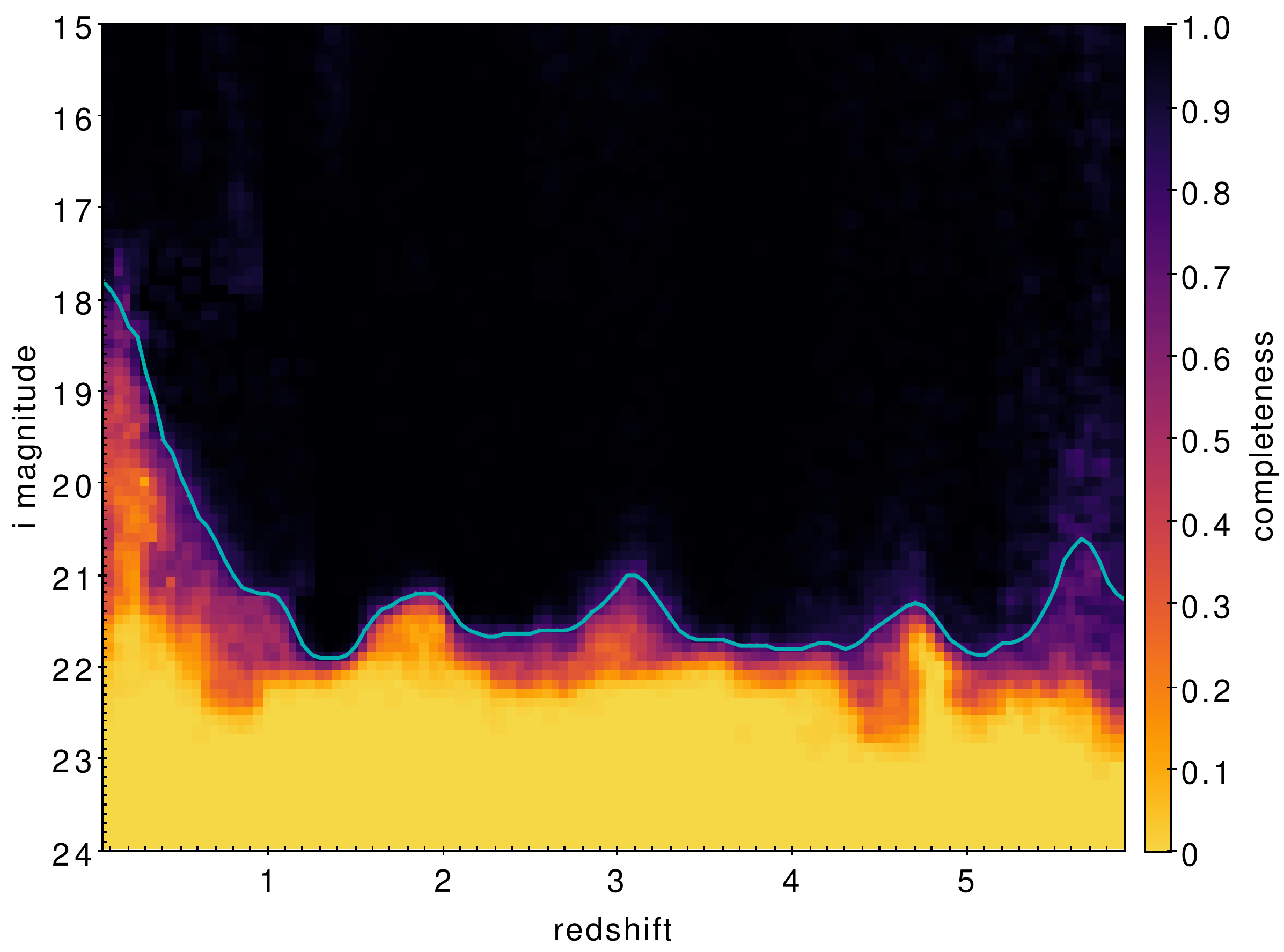}
\caption{2D selection completeness map as functions of redshift (x-axis) and $i$-band magnitude (y-axis) using criterion $P_{\rm QSO} > P_{\rm Galaxy}$ and $P_{\rm QSO} > P_{\rm Star}$ on a sample of simulated quasars. The color indicates the completeness. The solid cyan line shows the location where the completeness is 80\%.
\label{fig:completeness}}
\end{figure}

We use simulated quasars to remedy for the small sample size of real quasars at high redshift. The simulation procedure is described in \S\ref{sec:traning}. We quantify the selection completeness using a sample of $\sim$0.8 million simulated quasars, spanning a wide range of redshifts and magnitudes.
Figure \ref{fig:completeness} shows the 2D completeness as a function of redshift (x-axis) and $i$-band magnitude (y-axis) using $P_{\rm QSO} > P_{\rm Galaxy}$ and $P_{\rm QSO} > P_{\rm Star}$, color-coded by selection completeness. The solid cyan line shows the location where the completeness is $\sim$80\%. Figure \ref{fig:completeness} confirms that at low redshift ($z<1$), the completeness decreases with decreasing redshift due to increasing host galaxy contamination. At $z>3.3$, the completeness does not decrease with redshift, indicating that the trend observed in Figure \ref{fig:mag_z} is mainly due to the small sample statistics at high redshift. At $z>1$, the completeness is roughly constant and starts to decrease with increasing magnitude around $i\sim 21-22$. At certain redshifts, for example, $z\sim$1.8, 3.0, 4.8, and 5.5, the 80\% selection completeness is achieved at shallower magnitude limit due to contamination from different types of stars, with decreasing effective temperatures.

\subsection{Comparison to Gaia QSOC Redshifts}

\begin{figure*}
 \subfigure{
  \includegraphics[width=3.6in]{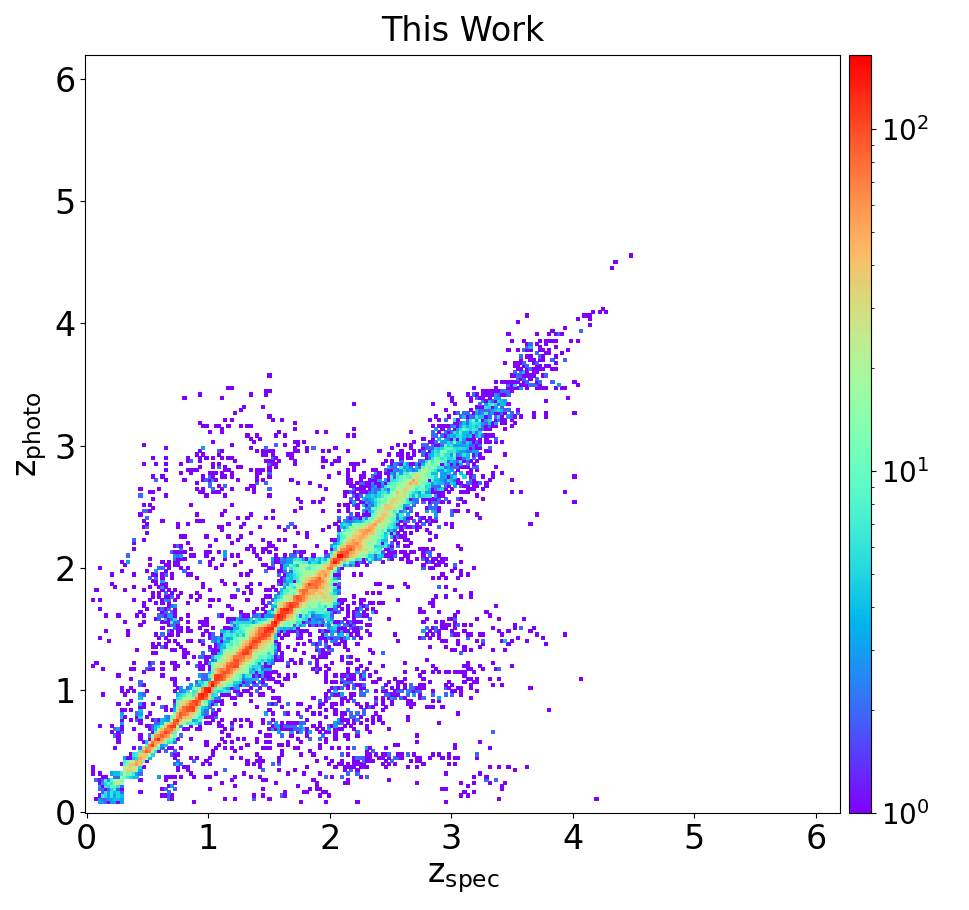}}
 \subfigure{
 \hspace{-0.5cm}
  \includegraphics[width=3.6in]{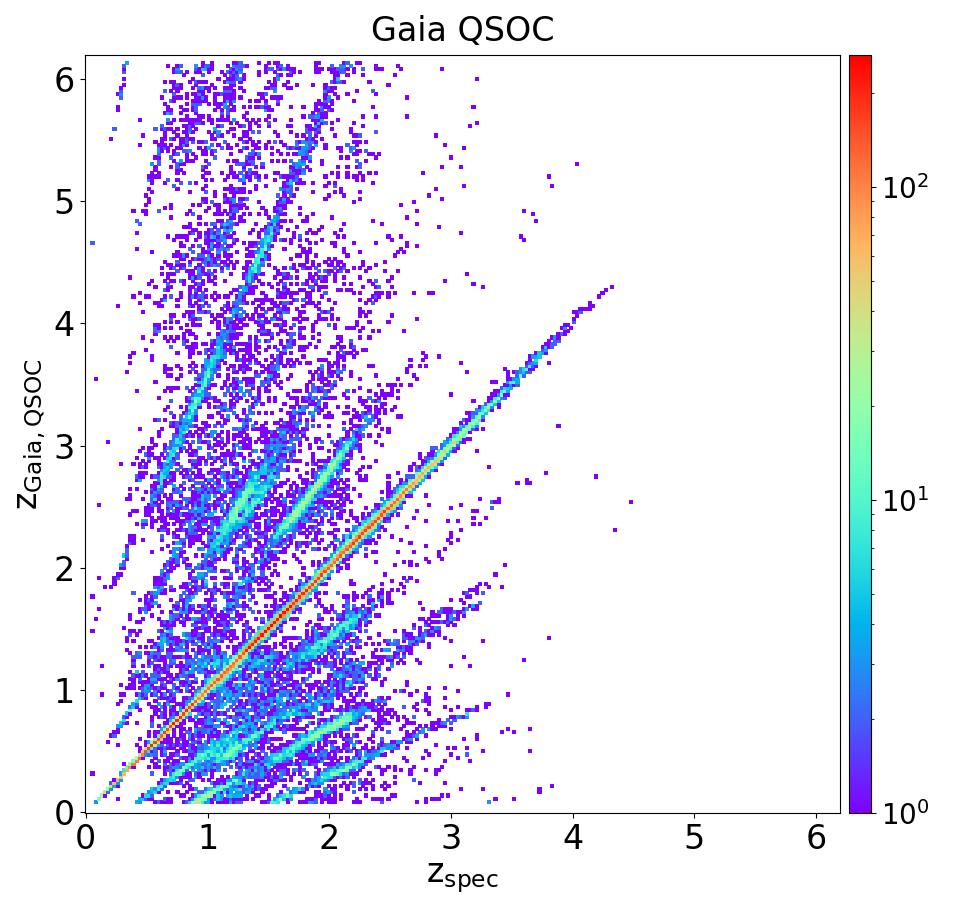}}\\
\caption{Comparison with the spectroscopic redshifts for our photozs (left panel) and Gaia QSOC low-spec redshifts (right panel), for 47451 spectrocopically confirmed quasars covered both in our DES photometric quasar sample and in Gaia QSOC. The colormap indicates the source number at each grid point. The photo-$z$ accuracy $R_{0.1}$ is 93.4\% and 61.1\% for our photo-zs and for Gaia QSOC redshifts, respectively. Thus the catastrophic redshift failure rate is much higher for the Gaia QSOC. 
\label{fig:DES_Gaia}}
\end{figure*}

Gaia DR3 release includes redshift estimates for extragalactic sources using low resolution BP/RP spectra \citep{Gaia2022}.
In Gaia DR3, the Quasi Stellar Object Classifier (QSOC) systematically publish redshift predictions for 1,834,118 sources, with a very low threshold on the Discrete Source Classifier (DSC) quasar probability of $classprob\_dsc\_combmod\_quasar \geq 0.01$ and a  warning flag of redshift estimation of $flags\_qsoc \leq 16$ \citep{Delchambre2022}.
There are 47451 spectroscopically confirmed quasars both in our 1.35 million quasar candidate catalog and in the 1.8 million Gaia QSOC redshift sample.
Figure \ref{fig:DES_Gaia} shows the comparison between photo-$z$s and spectroscopic redshifts for our algorithm (left panel) and the Gaia QSOC redshifts (right panel) for these 47451 quasars. The photo-$z$ accuracy $R_{0.1}$ is 93.4\% and 61.1\% from our algorithm and Gaia QSOC, respectively. 
Using photo-$z$s from our algorithm, the vast majority is along the 1:1 line.
Gaia QSOC redshifts from the low-resolution spectra have smaller scatter along the 1:1 line, indicating smaller redshift uncertainties than our photo-zs (as expected); but there are additional stripes that represent misidentified emission lines in Gaia low-resolution spectra. In particular, 4\% of the Gaia QSOC redshifts are incorrectly predicted at $z>4.6$, grossly over-predicting the abundance of high-redshift quasars. Using a more stringent cut of $flags\_qsoc = 0$ with empty warning flags, described by \citet{Delchambre2022}, the Gaia QSOC redshift accuracy $R_{0.1}$ increases to 95.0\%, but the sample is downsized to only 20\%. In comparison, using a higher-quality cut in our algorithm of $P\_{\rm QSO}\_z > 0.5$, i.e., the integrated probability of the identified photo-$z$ peak is higher than 0.5, our photo-$z$ accuracy $R_{0.1}$ increases to 94.7\% while the sample is only slightly downsized to 95\%.

\section{Summary} \label{sec:summary}
We perform quasar target selection in the southern hemisphere over the $\sim 5000$\,deg$^2$ DES-wide survey area. We utilize public DES DR2 optical photometry and available NIR photometric data, from various surveys including VHS, VIKING, UHS, ULAS, and 2MASS. In the MIR, we use the all-sky unWISE photometric data. Our algorithm can efficiently classify sources into the categories quasars, galaxies, and stars, as well as derive photo-$z$ for quasar and galaxy candidates. 

Our algorithm can successfully classify 94.7\% of quasars, 99.3\% of galaxies, and 96.3\% of stars when all bands are available, benchmarked on spectroscopically confirmed samples. The classification and photo-$z$ success rate decrease when fewer bands are available. The quasar (galaxy) photo-$z$ accuracy $R_{0.1}$, the fraction of objects with $|\Delta z|\equiv |z_s-z_p|/(1+z_s)$ smaller than 0.1, is as high as 92.2\% (98.1\%) when all bands are available, and decreases to 72.2\% (90.0\%) when only using 5-band photometry from DES. 

We select 1.4 million quasar candidates over the DES wide survey footprint, and provide all classification probabilities to customarily select quasar samples with different completeness and efficiency (purity). Selection completeness and efficiency are anti-correlated. We recommend to use our fiducial criteria (\S\ref{sec:selection}) for the most complete quasar sample. We recommend to use one more criterion of $P_{\rm QSO} > 0.7$ for a higher-purity selection and simultaneous high completeness  ($\sim85\%$). 

We provide our quasar, galaxy, and star probabilities for all $\sim 0.69$ billion sources in the DES DR2 coadd photometric catalog. This catalog will be useful for a broad range of extragalactic and galactic sciences in the southern hemisphere. 

\acknowledgments

QY and YS acknowledge support from NSF grants AST-1715579 and AST-2009947. 

Funding for the DES Projects has been provided by the U.S. Department of Energy, the U.S. National Science Foundation, the Ministry of Science and Education of Spain, the Science and Technology Facilities Council of the United Kingdom, the Higher Education Funding Council for England, the National Center for Supercomputing Applications at the University of Illinois at Urbana-Champaign, the Kavli Institute of Cosmological Physics at the University of Chicago, 
the Center for Cosmology and Astro-Particle Physics at the Ohio State University,
the Mitchell Institute for Fundamental Physics and Astronomy at Texas A\&M University, Financiadora de Estudos e Projetos, Funda{\c c}{\~a}o Carlos Chagas Filho de Amparo {\`a} Pesquisa do Estado do Rio de Janeiro, Conselho Nacional de Desenvolvimento Cient{\'i}fico e Tecnol{\'o}gico and the Minist{\'e}rio da Ci{\^e}ncia, Tecnologia e Inova{\c c}{\~a}o, the Deutsche Forschungsgemeinschaft and the Collaborating Institutions in the Dark Energy Survey. 

The Collaborating Institutions are Argonne National Laboratory, the University of California at Santa Cruz, the University of Cambridge, Centro de Investigaciones Energ{\'e}ticas, Medioambientales y Tecnol{\'o}gicas-Madrid, the University of Chicago, University College London, the DES-Brazil Consortium, the University of Edinburgh, the Eidgen{\"o}ssische Technische Hochschule (ETH) Z{\"u}rich, 
Fermi National Accelerator Laboratory, the University of Illinois at Urbana-Champaign, the Institut de Ci{\`e}ncies de l'Espai (IEEC/CSIC), 
the Institut de F{\'i}sica d'Altes Energies, Lawrence Berkeley National Laboratory, the Ludwig-Maximilians Universit{\"a}t M{\"u}nchen and the associated Excellence Cluster Universe, the University of Michigan, NSF's NOIRLab, the University of Nottingham, The Ohio State University, the University of Pennsylvania, the University of Portsmouth, SLAC National Accelerator Laboratory, Stanford University, the University of Sussex, Texas A\&M University, and the OzDES Membership Consortium.

Based in part on observations at Cerro Tololo Inter-American Observatory at NSF's NOIRLab (NOIRLab Prop. ID 2012B-0001; PI: J. Frieman), which is managed by the Association of Universities for Research in Astronomy (AURA) under a cooperative agreement with the National Science Foundation.

The DES data management system is supported by the National Science Foundation under Grant Numbers AST-1138766 and AST-1536171.
The DES participants from Spanish institutions are partially supported by MICINN under grants ESP2017-89838, PGC2018-094773, PGC2018-102021, SEV-2016-0588, SEV-2016-0597, and MDM-2015-0509, some of which include ERDF funds from the European Union. IFAE is partially funded by the CERCA program of the Generalitat de Catalunya. Research leading to these results has received funding from the European Research Council under the European Union's Seventh Framework Program (FP7/2007-2013) including ERC grant agreements 240672, 291329, and 306478. We acknowledge support from the Brazilian Instituto Nacional de Ci\^encia
e Tecnologia (INCT) do e-Universo (CNPq grant 465376/2014-2).

We acknowledge the use of SDSS data. Funding for SDSS-III has been provided by the Alfred P. Sloan Foundation, the Participating Institutions, the National Science Foundation, and the U.S. Department of Energy Office of Science. The SDSS-III website is \url{http://www.sdss3.org/}. SDSS-III is managed by the Astrophysical Research Consortium for the Participating Institutions of the SDSS-III Collaboration including the University of Arizona, the Brazilian Participation Group, Brookhaven National Laboratory, Carnegie Mellon University, University of Florida, the French Participation Group, the German Participation Group, Harvard University, the Instituto de Astrofisica de Canarias, the Michigan State/Notre Dame/JINA Participation Group, Johns Hopkins University, Lawrence Berkeley National Laboratory, Max Planck Institute for Astrophysics, Max Planck Institute for Extraterrestrial Physics, New Mexico State University, New York University, Ohio State University, Pennsylvania State University, University of Portsmouth, Princeton University, the Spanish Participation Group, University of Tokyo, University of Utah, Vanderbilt University, University of Virginia, University of Washington, and Yale University.

We acknowledge the use of LAMOST data. The Large Sky Area Multi-Object Fiber Spectroscopic Telescope (LAMOST, also named Guoshoujing Telescope) is a National Major Scientific Project built by the Chinese Academy of Sciences. Funding for the project has been provided by the National Development and Reform Commission. LAMOST is operated and managed by the National Astronomical Observatories, Chinese Academy of Sciences.

We acknowledge the use of NIR data, including VHS, VIKING, ULAS, and UHS. The UHS is a partnership between the UK STFC, The University of Hawaii, The University of Arizona, Lockheed Martin and NASA. This publication makes use of data products from the Two Micron All Sky Survey, which is a joint project of the University of Massachusetts and the Infrared Processing and Analysis Center/California Institute of Technology, funded by the National Aeronautics and Space Administration and the National Science Foundation. This publication makes use of data products from the \emph{Wide-field Infrared Survey Explorer}, which is a joint project of the University of California, Los Angeles, and the Jet Propulsion Laboratory/California Institute of Technology, funded by the National Aeronautics and Space Administration.

\appendix

\section{Quasar Targets for the Black Hole Mapper in SDSS-V}

We perform quasar target selection for the Black Hole Mapper (BHM) program in SDSS-V \citep{Kollmeier_etal_2017}, in particular, targets in the reverberation mapping (RM) fields. We use the same algorithm but utilize additional optical photometric data. Table \ref{tab:Fields} summarizes the photometric data we use for the seven initial BHM-RM fields. We use DES photometric data in the XMM-LSS, CDFS, EDFS, and ELAIS-S1 fields. We use the Pan-STARRS1 data \citep{Chambers2016} for the COSMOS and SDSS-RM fields in the northern sky. For the S-CVZ field we use optical photometric data from Gaia DR2 or the NOAO Source Catalog \citep[NSC; ][]{Nidever2018}. In addition, we make use of available NIR data in these fields. We use unWISE $W1$ and $W2$ photometric data in all fields. Since there are very few high-redshift ($i$-band dropouts at $z>5.8$) quasars in these small fields, we use $i$-band (or Gaia $G$-band) as the reference band for quasar target selection in these BHM-RM fields. The final quasar target catalog for BHM-RM fields is presented in Table~\ref{tab:catalog_bhm}. We required the Skewt-QSO probability criteria (i.e., $P\_{{\rm QSO}} > P\_{{\rm Star}}$ and $P\_{{\rm QSO}} > P\_{{\rm Galaxy}}$ in non-S-CVZ fields, and $P\_{{\rm QSO \_ Gaia}} > P\_{{\rm Star \_ Gaia}}$ and $P\_{{\rm QSO \_ Gaia}} > P\_{{\rm Galaxy \_ Gaia}}$ in the S-CVZ field). The SDSS-V BHM-RM quasar targets (v0.5) were selected from this catalog with further criteria on ${\rm log\_QSO}$, PLXSIG, PMSIG, and magnitude limits on $i$-band (or Gaia $G$-band) magnitude. Specifically, we used a criterion of ${\rm log\_QSO} > -10$ for SDSS-V BHM-RM quasar targets (v0.5). This criterion is explained in more detail in \citet{Yang2017}.

\section{A Catalog for All DES DR2 Sources}

We publicly release our quasar, galaxy, and star probabilities for all (0.69 billion) photometric sources in the DES DR2 coadded source catalog. We assign photoz and probability parameters as those of quasars when $P_{\rm QSO} > P_{\rm Galaxy}$, and as those of galaxies when $P_{\rm QSO} \leq P_{\rm Galaxy}$. The last three columns in Table~\ref{tab:catalog_all} are the probabilities that only use likelihood, i.e. no prior probabilities were used in Eq.~(\ref{eq:8}). These likelihood parameters are useful for redshift ranges where the luminosity functions may not be well measured, for example, for high-redshift quasars at $z>6$.

\begin{deluxetable*}{ccccc}
\tablecaption{Deep BHM-RM Fields in SDSS-V \label{tab:Fields}}
\tablewidth{1pt}
\tablehead{
\colhead{Field Name} &
\colhead{RA Center} &
\colhead{DEC Center} &
\colhead{Optical Survey} &
\colhead{Infrared Survey}
}
\startdata
XMM-LSS & 02:22:50.00 & $-$04:45:00.0 & DES & VHS \\
CDFS    & 03:30:35.60 & $-$28:06:00.0 & DES & VHS/VIKING/VIDEO \\
EDFS    & 04:04:57.84 & $-$48:25:22.8 & DES & VHS \\
ELAIS-S1 & 00:37:48.00  &  $-$44:00:00.0 & DES & VHS \\
COSMOS  & 10:00:00.00  &  $+$02:12:00.0 & PS1 & LAS \\
SDSS-RM & 14:14:49.00  &  $+$53:05:00.0 & PS1 & UHS \\
S-CVZ   & 06:00:00.00  &  $-$66:33:38.0 & Gaia/NSC & VHS/VMC \\
\enddata
\tablecomments{
VIDEO: VISTA Deep Extragalactic Observations Survey (DR5).\\
VMC: VISTA Magellanic Cloud Survey (DR4).
}
\end{deluxetable*}

\begin{deluxetable*}{llll}
\tablecaption{FITS Table Format for Quasar Targets for the BHM-RM Program in SDSS-V \label{tab:catalog_bhm}}
\tablewidth{1pt}
\tablehead{
\colhead{Column Name} &
\colhead{Format} &
\colhead{Units} &
\colhead{Description}
}
\startdata
% ------------ DES information ------------
FIELD\_NAME & STRING & & XMM-LSS, CDFS, EDFS, ELAIS-S1, COSMOS, SDSS-RM, or S-CVZ \\ 
POS\_REF & STRING & & Fiducial coordinates reference, priority: Gaia $>$ DES $>$ PS1 $>$ NSC \\
RA & DOUBLE & degree & Fiducial Right Ascension (J2000) \\
DEC & DOUBLE & degree & Fiducial Declination (J2000) \\
Distance & DOUBLE & degree & Angular distance from the field center \\
EBV & DOUBLE & mag & Galactic $E(B - V)$ reddening from \citet{Schlegel1998} \\
DES & INT & & A flag set to 1 if in DES photometric catalog \\
DES\_MAG & DOUBLE & mag & DES PSF magnitudes in $grizY$ bands\\
DES\_MAGERR & DOUBLE & mag & DES PSF magnitude uncertainties in $grizY$ bands\\
PS1 & INT & & A flag set to 1 if in PS1 photometric catalog \\
PS1\_MAG & DOUBLE & mag & PS1 PSF magnitudes in $grizy$ bands\\
PS1\_MAGERR & DOUBLE & mag & PS1 PSF magnitude uncertainties in $grizy$ bands\\
NSC & INT & & A flag set to 1 if in NSC catalog \\
NSC\_MAG & DOUBLE & mag & NSC PSF magnitudes in $grizY$ bands\\
NSC\_MAGERR & DOUBLE & mag & NSC PSF magnitude uncertainties in $grizY$ bands\\
SDSS & INT & & A flag set to 1 if in SDSS photometric catalog \\
SDSS\_MAG & DOUBLE & mag & SDSS PSF magnitudes in $ugriz$ bands\\
SDSS\_MAGERR & DOUBLE & mag & SDSS PSF magnitude uncertainties in $ugriz$ bands\\
Gaia & INT & & A flag set to 1 if in Gaia photometric catalog \\
Gaia\_MAG & DOUBLE & mag & Gaia magnitudes in $G$, $G_{BP}$, and $G_{RP}$ bands\\
Gaia\_MAGERR & DOUBLE & mag & Gaia magnitude uncertainties in $G$, $G_{BP}$, and $G_{RP}$ bands\\
% ------------ Gaia proper motion ------------
PLXSIG & FLOAT & & Gaia DR2 parallax significance\\
PMSIG & FLOAT & & Gaia DR2 proper motion significance\\
% ------------ WISE ------------
WISE & INT & & A flag set to 1 if in WISE photometric catalog \\
WISE\_MAG & DOUBLE & mag & WISE magnitudes in $W1$, and $W2$ bands (in Vega magnitude)\\
WISE\_MAGERR & DOUBLE & mag & WISE magnitude uncertainties in $W1$, and $W2$ bands\\
Separation\_WISE & FLOAT & arcsec & Angular distance between WISE and the fiducial coordinates  \\
NIR & INT & & A flag set to 1 if in NIR photometric catalog \\
Survey\_NIR & STRING & & NIR survey \\
NIR\_MAG & DOUBLE & mag & NIR magnitudes in $YJHK$ bands (in Vega magnitude)\\
NIR\_MAGERR & DOUBLE & mag & NIR magnitude uncertainties in $$YJHK$$ bands\\ 
Separation\_NIR & FLOAT & arcsec & Angular distance between NIR and the fiducial coordinates  \\
% ------------ Skewt-QSO ------------
Optical\_Survey & STRING & & Optical survey used in Skewt-QSO, e.g., DES, PS1, Gaia, NSC \\
mi & DOUBLE & mag & i-band PSF magnitude \\
Nband\_Optical & INT & & Number of optical bands used in Skewt-QSQ \\
Combination & STRING & & Optical, NIR, and MIR survey combination \\
log\_QSO & DOUBLE & & The natural logarithmic probability of a target fitting to QSO (Eq. \ref{eq:9}) \\
$P\_{\rm QSO}$ & FLOAT & & $P_{\rm QSO}$, Skewt-QSO probability fitting to QSO models, described in Eq. \ref{eq:10} \\
$P\_{\rm Galaxy}$ & FLOAT & & $P_{\rm Galaxy}$, Skewt-QSO probability fitting to Galaxy models \\
$P\_{\rm Star}$ & FLOAT & & $P_{\rm Star}$, Skewt-QSO probability fitting to Star models  \\
$P\_{\rm QSO}$\_Gaia & FLOAT & & $P\_{\rm QSO}$ using Gaia/NSC photometric data (for the S-CVZ field) \\
$P\_{\rm Galaxy}$\_Gaia & FLOAT & & $P\_{\rm Galaxy}$ using Gaia/NSC photometric data (for the S-CVZ field) \\
$P\_{\rm Star}$\_Gaia & FLOAT & & $P\_{\rm Star}$ using Gaia/NSC photometric data (for the S-CVZ field) \\
photoz\_QSO & FLOAT & & Quasar photo-$z$\\
$z1\_{\rm QSO}$ & FLOAT & & The lower limit of quasar photo-$z$ \\
$z2\_{\rm QSO}$ & FLOAT & & The upper limit of quasar photo-$z$ \\
$P\_{{\rm QSO}}\_z$ & DOUBLE & & The probability of quasar photo-$z$ locating within ($z1_{\rm QSO}$, $z2_{\rm QSO}$), described in Eq. \ref{eq:5}  \\
\enddata
\end{deluxetable*}

\begin{deluxetable*}{llll}
\tablecaption{FITS Table Format for All DES-DR2 Sources \label{tab:catalog_all}}
\tablewidth{1pt}
\tablehead{
\colhead{Column} &
\colhead{Format} &
\colhead{Units} &
\colhead{Description}
}
\startdata
% ------------ DES information ------------
COADD\_OBJECT\_ID	& LONG64 & & Unique identifier for the coadded objects \\       
ALPHAWIN\_J2000 & DOUBLE & degree & DES Right Ascension (J2000) \\
DELTAWIN\_J2000 & DOUBLE & degree & DES Declination (J2000) \\
EXTENDED\_COADD & INT & & DES morphological object classification variable \\
 & & & $~~~~$0: high confidence pointed-like; 1: likely pointed-like; \\
 & & & $~~~~$2: likely extended; 3: high confidence extended\\
SN\_MAX\_PSF & FLOAT & & MAX SNR of the PSF mag in DES \\
SN3 & INT & & The number of bands in DES with SNR higher than 3 \\
Photometry & STRING & & DES photometry fitting to quasar and star models, PSF or AUTO \\
Band\_DES & STRING & & DES bands \\
Reference\_Band & STRING & & DES reference band \\
IMAFLAGS\_ISO\_GRIZY & INT & & DES flag in $grizY$ bands \\
% ------------ Gaia proper motion ------------
PLXSIG & FLOAT & & Gaia parallax significance\\
PMSIG & FLOAT & & Gaia proper motion significance\\
Separation\_Gaia & FLOAT & arcsec & Angular distance between DES and Gaia coordinates \\
CNT9 & INT & & Number of sources with a 9 arcsec radius circular aperture \\
DIST & FLOAT & arcsec & Angular distance to the closest neighbour within 9 arcsec\\
% ------------ IR ------------
Survey\_NIR & STRING & & NIR survey \\
Band\_NIR & STRING & & NIR bands \\
Band\_WISE & STRING & & WISE bands (only use $W1$ and $W2$ bands)\\
Separation\_WISE & FLOAT & arcsec & Angular distance between DES and unWISE coordinates  \\
% ------------ Skewt-QSO ------------
Combination & STRING & & DES, NIR, and MIR band combination \\
Class & STRING & & The classification from Skewt-QSO probabilities: ``QSO", ``Galaxy", or ``Star" \\
$P\_{\rm Star}$ & FLOAT & & Same as $P_{\rm Star}$ in Table \ref{tab:catalog} \\ % Class\_Skewt\_QSO
\hline
Class\_photoz & STRING & & photoz class: ``QSO" when $P_{\rm QSO} > P_{\rm Galaxy}$; ``Galaxy"  when $P_{\rm QSO} \leq P_{\rm Galaxy}$   \\
$P$ & FLOAT & & $P_{\rm QSO}$ when $P_{\rm QSO} > P_{\rm Galaxy}$; $P_{\rm Galaxy}$ when $P_{\rm QSO} \leq P_{\rm Galaxy}$  \\
photoz & FLOAT & & photoz-QSO when $P_{\rm QSO} > P_{\rm Galaxy}$; photoz-Galaxy when $P_{\rm QSO} \leq P_{\rm Galaxy}$   \\
$z1$ & FLOAT & & $z1_{\rm QSO}$ when $P_{\rm QSO} > P_{\rm Galaxy}$; $z1_{\rm Galaxy}$ when $P_{\rm QSO} \leq P_{\rm Galaxy}$ \\
$z2$ & FLOAT & & $z2_{\rm QSO}$ when $P_{\rm QSO} > P_{\rm Galaxy}$; $z2_{\rm Galaxy}$ when $P_{\rm QSO} \leq P_{\rm Galaxy}$ \\
$P\_{z}$ & FLOAT & & $P_{{\rm QSO}z}$ when $P_{\rm QSO} > P_{\rm Galaxy}$; $P_{{\rm Galaxy}z}$ when $P_{\rm QSO} \leq P_{\rm Galaxy}$  \\
\hline
Class\_other & STRING & & the other class: ``Galaxy" when $P_{\rm QSO} > P_{\rm Galaxy}$; ``QSO"  when $P_{\rm QSO} \leq P_{\rm Galaxy}$   \\
$P$\_other & FLOAT & & $P_{\rm Galaxy}$ when $P_{\rm QSO} > P_{\rm Galaxy}$; $P_{\rm QSO}$ when $P_{\rm QSO} \leq P_{\rm Galaxy}$  \\
photoz\_other & FLOAT & & photoz-Galaxy when $P_{\rm QSO} > P_{\rm Galaxy}$; photoz-QSO when $P_{\rm QSO} \leq P_{\rm Galaxy}$   \\
$z1$\_other & FLOAT & & $z1_{\rm Galaxy}$ when $P_{\rm QSO} > P_{\rm Galaxy}$; $z1_{\rm QSO}$ when $P_{\rm QSO} \leq P_{\rm Galaxy}$ \\
$z2$\_other & FLOAT & & $z2_{\rm Galaxy}$ when $P_{\rm QSO} > P_{\rm Galaxy}$; $z2_{\rm QSO}$ when $P_{\rm QSO} \leq P_{\rm Galaxy}$ \\
$P\_{z}$\_other & FLOAT & & $P_{{\rm Galaxy}z}$ when $P_{\rm QSO} > P_{\rm Galaxy}$; $P_{{\rm QSO}z}$ when $P_{\rm QSO} \leq P_{\rm Galaxy}$  \\
\hline
$P{\_\rm QSO\_likelihood}$ & FLOAT & & $P_{\rm QSO}$ that only use likelihood\\
$P{\_\rm Galaxy\_likelihood}$ & FLOAT & & $P_{\rm Galaxy}$ that only use likelihood\\
$P{\_\rm Star\_likelihood}$ & FLOAT & & $P_{\rm Star}$ that only use likelihood\\
\enddata
\end{deluxetable*}

\bibliography{extracted.bib}

\end{document}